\documentclass{aa}
\usepackage{graphicx,times}

\font\cal=cmsy10
\newcount\lecurrentfam
\def\LaTeX{\lecurrentfam=\the\fam \leavevmode L\raise.42ex
\hbox{$\fam\lecurrentfam\scriptstyle\kern-.3em A$}\kern-.15em\TeX}

\begin{document}
%
%
%
%
\title{The gradient of diffuse $\gamma$-ray emission in the Galaxy}

%
%
\authorrunning{D.\ Breitschwerdt et al.}
\titlerunning{The gradient of diffuse $\gamma$-ray emission in the Galaxy}
%
\author {D.\ Breitschwerdt\inst{1}, V.A.\ Dogiel\inst{2,}\inst{3},
and H.J.\ V\"olk\inst{4}}
\offprints{D.\ Breitschwerdt}

\institute{
Max-Planck Institut f\"ur extraterrestrische Physik, Postfach 1312, 
D-85741 Garching, Germany \\
   \email{breitsch@mpe.mpg.de}
\and
P.N.Lebedev Physical Institute, 119991 Moscow GSP-1, Russia \\
   \email{dogiel@lpi.ru}
\and
Institute of Space and Astronautical Science, 
3-1-1 Yoshinodai, Sagamihara, Kanagawa 229-8510, Japan \\
   \email{vad@astro.isas.ac.jp}
\and
Max-Planck Institut f\"ur Kernphysik, Postfach 103 980,
D-69029 Heidelberg, Germany \\
   \email{Heinrich.Voelk@mpi-hd.mpg.de}
  }%

\date{Received 12 June, 2001; accepted 16 January, 2002}

\abstract{
We show that the well-known discrepancy, known for about two decades,
between the radial dependence of the Galactic cosmic ray nucleon 
distribution, as inferred
most recently from EGRET observations of diffuse $\gamma$-rays above 100
MeV, and of the most likely cosmic ray source distribution (supernova 
remnants,
superbubbles, pulsars)  can be explained purely by {\em propagation}
effects. Contrary to previous claims, we demonstrate that this is possible,
if the dynamical coupling between the escaping cosmic rays and
the thermal plasma  is  taken into account, and thus a self-consistent
calculation of a {\em Galactic Wind} is carried out. Given a dependence of 
the cosmic ray source distribution on Galactocentric radius  $r$, our
numerical wind solutions show that the cosmic ray outflow
velocity, $V(r,z) = u_0 + V_{{\rm A} 0}$, also depends both on
$r$, as well as on  vertical distance $z$, with
$u_0$ and $V_{{\rm A} 0}$ denoting the thermal gas and the
Alfv\'en velocities, respectively, at a reference level $z_{\rm C}$. The
latter is by definition the transition boundary from diffusion to advection
dominated cosmic ray transport and is therefore also a function of $r$.  
In fact,  the cosmic ray escape time averaged over particle energies
decreases with increasing cosmic ray source strength.
Thus an increase in cosmic ray source strength is counteracted by a reduced 
average
cosmic ray residence time in the gas disk. This means that pronounced 
peaks in the radial distribution of the source strength
result  in  mild radial $\gamma$-ray gradients at GeV energies,
as it has been observed. The effect might be enhanced by anisotropic
diffusion, assuming different radial and vertical diffusion coefficients.
In order to better understand the mechanism described, we have calculated 
analytic solutions of the stationary diffusion-advection equation, including
anisotropic diffusion in an axisymmetric geometry, for a given
cosmic ray source distribution and a realistic outflow velocity field
$V(r,z)$, as inferred from the self-consistent numerical Galactic Wind
simulations performed simultaneously. At TeV energies the $\gamma$-rays 
from the sources themselves are expected to dominate the observed 
``diffuse'' flux from the disk. Its observation should therefore allow an 
empirical test of the theory presented. 
}

\maketitle

%

\section{Introduction}

Our information on the spatial distribution of cosmic rays (CRs) in the
Galaxy stems largely from measurements of nonthermal emission, generated by
the energetic charged particles interacting with matter and 
electromagnetic
fields. For $\gamma$-ray energies above
100 MeV, the main production process is probably via $\pi^0$-decay,
resulting from nuclear collisions between high energy particles and
interstellar matter.  Past and recent observations in the GeV range have 
shown a roughly
uniform distribution of diffuse $\gamma$-ray emissivity in the Galactic plane,
exhibiting only a shallow radial gradient (in a cylindrical Coordinate 
system).
Hence, if $\gamma$-rays were to map the spatial CR distribution, we would
expect it to be uniform as well. However, associating CR production regions 
with star formation regions, all possible Galactic CR {\em source} 
distributions are strongly peaked towards a Galactocentric distance at which
a ring of molecular gas resides.
It is commonly believed that the bulk of the CR nucleons below about
$10^{15} \, {\rm eV}$ is produced in supernova remnants (SNRs) - the majority  
being core collapse SNRs - most likely by 
diffusive shock acceleration (Krymsky 1977; Axford et al. 1977; Bell
1978a,b;  Blandford  \& Ostriker 1978).
It has been  argued that up
to some tenths of the hydrodynamic explosion energy might be 
converted  into CR energy  (e.g., Berezhko et al.
1994).  Since high-mass star formation mostly occurs in a spatially
nonuniform manner, i.e.\ in OB associations predominantly located in the
spiral arms of late type galaxies, we are confronted with the problem of
reproducing a mild radial gradient in the diffuse Galactic $\gamma$-ray
emission as it has been observed for the first time by the COS-B satellite
(Strong et al. 1988) and, more recently, with higher angular and energy
resolution, higher sensitivity and lower background by the EGRET instrument
of the CGRO satellite (Strong \& Mattox 1996, Digel et al. 1996). If the
SNRs are the sources of the CR nucleon component and if
the source distribution is inhomogeneous, this discrepancy must arise during
the {\em propagation} of CRs from their sources through the interstellar
medium.
Unlike e.g.\ the interpretation of radio synchrotron emission, generated by
relativistic electrons, $\gamma$-ray data open the 
possibility of studying the nucleonic component of the CRs, in which almost
all of the energy is stored (see e.g., Dogiel \& Sch\"onfelder 1997).
The distribution of the $\gamma$-ray emissivity in the Galactic disk
therefore bears important information on the  origin of CRs and on the
conditions of CR propagation in the Galaxy.

\section{Gradient of the $\gamma$-Ray Emissivity in the Galactic Disk} 
The first data on the radial distribution (i.e.\ gradient) of the
$\gamma$-ray emissivity in the disk were obtained with the SAS-II
satellite (energy range 30 -- 200 MeV). The data showed rather strong
variations of the emissivity along the Galactic plane which dropped
rapidly with radius (see e.g., Stecker \& Jones 1977), in rough agreement
with the distribution of candidate CR sources in the Galaxy such as
SNRs or pulsars. A noticeable discrepancy however emerged from the
COS-B data (energy range 70 -- 5000 MeV), in which the emissivity gradient
was found to be rather small, when compared to the SNR distribution,
especially at high energies ($E_\gamma >$300 MeV), with a maximum
variation by a factor of only 2 (see Strong et al. 1988). An energy
dependence of the gradient was mentioned by almost all groups analyzing
the COS-B data and was usually interpreted as a steeper gradient for CR
electrons, producing the soft part of the $\gamma$-ray spectrum, compared
to nuclei. Detailed analyses of different models of CR propagation based
on the gradient value were usually performed with the COS-B data because
of their much better statistics compared with the SAS-II data.

The recent measurements with the EGRET instrument onboard CGRO, obtained
by different methods at energies 100 -- 10000 MeV, showed that the
emissivity drops at the edge of the disk. Digel et al. (1996), performed a
study of the outer part of the Galaxy towards molecular complexes, the
Cepheus and Polaris Flares in the local arm and the larger molecular
complex in the Perseus arm. Since the total masses of these complexes are
known, it was possible to infer the CR density from the measured
$\gamma$-ray fluxes in the direction of these objects. It was discovered
that the apparent emissivity decreases by a factor 1.7, which is somewhat
smaller 
than that for COS-B. The analysis of Strong \& Mattox (1996), based on a
model of the average gas density distribution in the disk, shows a smaller
intensity gradient which does not differ significantly from that of COS-B
in the Galactic disk.

One of the important conclusions which follows from all these data, and
which we will discuss below, is that the distribution of the 
$\gamma$-emissivity in the disk in the GeV range is rather uniform
compared with the most probable distribution of CR sources.

\section{Basic Ideas for a New Approach of CR Propagation in the Galaxy  }

A natural explanation of a uniform CR distribution would be effective
radial mixing due to the diffusion of CRs produced in different
parts of the Galactic disk. It is then straightforward to infer the mixing
volume, which usually includes the Galactic disk plus a large Galactic halo;
the details of such a model were discussed in the book of Ginzburg and
Syrovatskii (1964). This 3-dimensional cylindrical model, which includes
diffusive-advective transport,  
a free escape of cosmic rays from the halo boundary into
intergalactic space, and the observed supernova shells as sources of 
cosmic
rays in the Galactic disk, explained very well characteristics of the
Galactic radio and $\gamma$-ray emission as well as the data on cosmic ray
spectra and their chemical composition including stable and radioactive
secondary nuclei, intensities of positrons and antiprotons etc. (a summary
of this analysis can be found in Berezinskii et al. 1990). 

In general the propagation equation for CRs is described by 
\begin{equation}
-\nabla_i (D_{ij}\nabla_j n - U_i n)+{{\partial}\over {\partial E}}
\left({{dE}\over{dt}}n\right)=Q(r,E) \,, \label{diff-trans1}
\end{equation}
where $D_{ij}$ is the diffusion tensor and $U_i$ is the advection velocity, 
which in general are functions of the coordinates, and $dE/dt$ is the term which 
describes particle energy losses including adiabatic losses if the velocities 
vary spatially. The right hand side of the equation describes the spatial 
and energy 
dependence of the source density, derived from radio observations of the 
SNR distribution and the injection spectrum of CRs.

The main
conclusion was that almost all observations can be reasonably explained if
the halo extension is about several kpc, the injection spectrum of electrons
and protons $Q(E)$ is a power-law ($Q(E)\propto E^{-\gamma_0}$ with
$\gamma_0\simeq 2.2 - 2.4$), and the radially uniform velocity of advection
does not exceed the value of 20 km/s. Compton scattering of relativistic
electrons was found to play a significant r\^ole in the halo $\gamma$-ray
emission. However, this model failed to explain a smooth emissivity
distribution of $\gamma$-rays in the Galactic disk (see for details Dogiel
\& Uryson 1988). Even in the case of a very extended halo with a radius
larger than 10 kpc the derived emissivity gradient (calculated for the
observed SN distribution) was larger than  observed. Only for a
hypothetical uniform distribution of the sources in the disk the
calculations can reproduce the data.

This model was further developed by Bloemen et al. (1991, 1993). Extensive
investigations of the 3-dimensional diffusion-advection transport equation
for nucleons, low-energy electrons and the $\gamma$-rays showed that even in
the most favorable case of an extended halo with a vertical height as large
as 20 kpc, their model, although reproducing the COS B data marginally, is
not able to remove the signature of the observationally inferred SNR 
source distribution, i.e. the distribution of the calculated
emissivity was still steeper than permitted by the data. The vertical
gradient of the advection velocity derived from the fluxes of stable and
radioactive nuclei near Earth had to be smaller than 15 km/s/kpc. The other 
unexpected conclusion was that the halo extension obtained from
the nuclear data was significantly less than estimated from the $\gamma$-ray
data (see also Webber et al. 1992).

Recently, Strong and collaborators in a series of papers (see Strong et al. 
2000, and references therein) have developed a numerical method
for this model and made a new attempt to analyze the $\gamma$-ray emission
and the CR data, based on the latest data from COMPTEL, EGRET and OSSE.
They limited the analysis to radially uniform diffusion-advection and to
diffusion-reacceleration transport models, and concluded that no such
diffusion-advection model can adequately describe the data, in particular
the B/C ratio and the energy dependence.  These authors claimed, however,
that by including reacceleration one can account for all the 
observational data. A peculiar
consequence of their analysis was that they did not use the {\em observed}
supernova source distribution (see e.g., Kodeira 1974, Leahy \& Xinji 1989,
Case \& Bhattacharya 1996, 1998) as the {\em input} $Q$ for Eq. (1), but
rather {\em derived} it from the $\gamma$-ray data to reproduce the observed
spatial variations of the emissivity in the disk. The result was a source
distribution that is flat in the radial direction.

Thus we see that the conventional model of unifrom diffusion-advection has
serious problems in spite of its evident achievements. One solution is to
assume that some of the observational data are not significant like the SN
distribution derived from the radio observations (Strong et al. 2000). 
The alternative is to conclude that it is time to abandon the
standard model, which is what we do in this paper. We shall demonstrate,
that strong radial source gradients will be removed by a strong {\em
advection} velocity in the halo (due to a Galactic wind driven by the CRs
themselves, see below) that varies with radius $R$ and height $z$. 
In addition {\em 
anisotropic diffusion} with different diffusion coefficients $D_\parallel$
and $D_\perp$ in the disk and the halo, respectively, might also play a
r\^ole. It should be emphasized that a radially varying advection velocity
occurs naturally in spiral galaxies, even for a uniform source
distribution, because the gravitational potential increases towards the
centre, thus inducing stronger velocity gradients in this direction
(see Fig.~\ref{snd1}), as has been shown by Breitschwerdt et al. (1991).

The existence of strong advective CR transport in the Galactic halo has 
been shown
on dynamical grounds in a number of papers in the past
(Ipavich 1975; Breitschwerdt et al. 1987, 1991, 1993;  Fichtner et al. 1991;
Zirakashvili et al. 1996; Ptuskin et al. 1997). The key element of halo 
transport theory is that
CRs, which by observations are known to escape from the Galaxy, resonantly
generate waves by the so-called streaming instability (Kulsrud \& Pearce 1969) 
leading to strong scattering of CRs. Therefore,
even in the case of strong non-linear  wave damping, advection is 
at least as important a CR transport mechanism out to large distances in 
the halo as diffusion, 
provided that the level of MHD-turbulence is high enough for coupling 
between CRs and MHD waves (Dogiel et al. 1994, Ptuskin et al. 1997).
There is also growing indirect observational evidence of outflows from the
interpretation of soft X-ray data of galactic halos in edge-on galaxies like
\object{NGC 4631} (Wang et al. 1995), and also of the soft X-ray background 
in our own
Galaxy (Breit\-schwerdt \& Schmutzler 1994, 1999). Furthermore, the 
near constancy of the spectral index of nonthermal radio continuum emission 
over large distances along the minor axis in
the halo of edge-on galaxies is most naturally explained by an advective
transport velocity of relativistic electrons (along with the nucleons) that
is ever increasing with distance from the Galactic plane (Breitschwerdt
1994).

In the disk of spiral galaxies, the regular magnetic field is following
roughly the spiral arms and is therefore mostly parallel to the disk, with
noticeable deviations in some regions where outflow is expected.
Here, also a regular vertical component seems to be present (eg.\ Hummel 
et al. 1988), 
which has been detected in a number of galaxies like \object{NGC 4631}, 
\object{NGC 5775} (T\"ullman et al. 2000), and \object{NGC 4217}. 
Multi-wavelength observations of the galaxy \object{NGC 253} show a local
correlation between non-thermal radio continuum, H$\alpha$ and X-ray emission
near the disk-halo interface in off-nuclear regions (Dettmar 1992; M. Ehle,
private communication). This also spatially coincides with enhanced star
formation activity in the disk as can be seen from FIR data.

Since the disk is not fully ionized in contrast to the halo and since waves
are efficiently dissipated there by ion-neutral damping, the most 
important
contribution to the random field in the disk is by turbulent mass motions,
induced by supernova explosions and other stellar mass loss activity. Thus 
the wave spectrum will be very
different from the one in the halo, where self-excited waves subject to
nonlinear wave damping (Dogiel et al. 1994, Ptuskin et al.  1997),
satisfying the gyro-resonance condition, dominate. Consequently, the
averaged diffusion coefficient will be different in the Galactic disk and
the halo (see Sect.~\ref{anid}). Therefore, we believe that anisotropic 
diffusion, together with
radially varying advection, is the most general and most probable mode of CR
transport to occur.
    
We investigate the CR transport processes of diffusion and advection and
discuss the possibility of flattening radial CR source gradients of a {\em
given SNR distribution} (despite observational uncertainties) by
particle propagation. In our view it is using the ``natural'' boundary
condition when {\em calculating} the response of CR transport to a given
CR source distribution (whatever its observational limitations may be at
the present time).  This is in contrast to {\em adjusting} the source
distribution a posteriori.  To implement the transport processes properly
we shall allow for radially varying advection, anisotropic diffusion,
(different values of the diffusion coefficient parallel and perpendicular
to the disk) and the appropriate boundary conditions to employ a cosmic
ray distribution and then to calculate from these the resulting spatial
distribution of CRs.

In Sect.~\ref{anid} we discuss the question of anisotropic CR diffusion
and in Sect.~\ref{sources} we introduce a simple model that describes
the {\em observed} CR source distribution. In the following
Sect.~\ref{flagra} we show heuristically (and in Sect.~\ref{crtrans}
also analytically)  how a radially varying CR source distribution induces
variations in the CR energy density, which in turn leads to a radial
variation of the diffusion-advection boundary, $z_{\rm C}$, and the outflow
velocity, respectively, and thus to a tendency to {\em flatten} the radial
CR distribution and hence the $\gamma$-ray emissivity gradient. In
Sect.~\ref{radvar} we demonstrate by numerical calculations how such a
source distribution will naturally generate a radially varying outflow
velocity.  In Sect.~\ref{crtrans} we discuss in detail models of
different complexity for advective-diffusive transport with radially
varying outflow velocity, and show analytically how in each case for a
{\em given} CR source distribution its radial signature on the energetic
particle and $\gamma$-ray distribution is reduced. The most advanced of
these models include 
a dependence of galactic wind velocity on the CR source strength. 
In addition a full analytic solution of the
3-dimensional diffusion-advection equation in axisymmetry for a given
realistic velocity dependence $u(r,z)$, parallel and perpendicular to the
Galactic disk similar to the one derived in Sect.~\ref{radvar} is 
calculated. In
Sect.~\ref{sourcecon} the source contributions to the ``diffuse''
$\gamma$-ray background at TeV energies are taken into account and shown
to be highly significant. This should allow an empirical test of our
theoretical picture.  In Sect.~\ref{discon} we discuss and summarize our
results. A number of detailed calculations can be found in the
Appendices~\ref{appa}, \ref{appb} and \ref{appc}.

\section{Anisotropic Diffusion Coefficient}
\label{anid}
In most models of diffusive CR propagation, the diffusion tensor is
approximated by a scalar quantity $D$, representing spatially uniform 
transport,
$D=D_0 E^\alpha$, where $\alpha=0.5 - 0.6$ describes the energy dependence.
However, there are good reasons why the diffusion coefficient may be
anisotropic.

The propagation of CRs in the interstellar medium is mainly determined by
their interaction with electrical and magnetic fields. CRs interact strongly
with fluctuations of the magnetic field (MHD-waves) and are scattered
by them in pitch angle.
In the simplest case the total magnetic field consists of two components
and can be written as
\begin{equation}
B=B_0+\delta B\,,
\end{equation}
where $B_0$ is the average large scale magnetic field and $\delta B$ is 
the
field of small scale fluctuations.

Effective scattering of  particles by these fluctuations occurs when the
interaction is resonant, i.e.
when the scale of the fluctuations in the magnetic field $B_0$ is of the
order of the particle gyroradius.
This leads to a stochastic motion of particles through space; 
the associated diffusion coefficient $D_\|$ along
the magnetic field $B_0$ is of the order of
\begin{equation}
D_\| \sim v L_\|\,,
\end{equation}
where $v$ is the
particle velocity. The value $L_\| \sim v/\nu$ is the transport length of 
the particle
along the magnetic field lines, and $\nu$ is the rate of resonant particle 
scattering. The main assumption of this
description is that the fluctuation amplitude is much less 
than the
average magnetic field
\begin{equation}
\delta B\ll B_0\,.
\end{equation}
The displacement of a charged particle perpendicular to the regular magnetic
field $B_0$ due to scattering is therefore small:
\begin{equation}
D_\perp \sim D_\|\cdot
\left({{r_{\rm g}}\over L_\|}\right)^2\,,
\label{d_larm}
\end{equation}
i.e.,
\begin{equation}
D_\| \gg D_\perp \,,
\end{equation}
where $r_{\rm g}$ is the particle gyroradius, and $r_{\rm g}\ll L_\|$.

The situation is more complex if there exists also a large scale random 
magnetic field $\Delta B$
whose scale is much larger than the particle gyroradius:
\begin{equation}
B=B_0+\Delta B +\delta B\,.
\end{equation}
In this case the diffusion transverse to the magnetic field $B_0$ occurs
due to the divergence of the particle trajectories, which is much faster than
the resonant diffusion from Eq.~\ref{d_larm}.

The procedure to derive the transport equation for CRs in this case was
described , e.g. by Toptygin (1985), who showed that the maximum value of
$D_\perp$ is:
\begin{equation}
D_\perp\sim D_\|\left({{\Delta B}\over {B_0}}\right)^2\,.
\end{equation}
Thus, diffusion in the Galaxy is quasi-isotropic 
only in the
case $\Delta B \sim B_0$.

A more precise analysis (see Berezinsky et al. 1990, chap. 9) shows,
however, that in the interstellar medium the correlation between the
components of the diffusion tensor leads to an effective perpendicular 
diffusion coefficient:
\begin{equation}
D_\perp^{\rm eff} \sim 0.1\cdot  D_\|\,.
\end{equation}

Observations show that in the
Galactic disk the large scale magnetic field is typically parallel to the
disk, whereas in the halo also a significant perpendicular
component may exist, as it has been observed in the case of the 
edge-on Sc galaxy \object{NGC 4631}. Therefore one can expect
$D_{\rm z}\ll \kappa_{\rm r}$ in the disk, and $D_{\rm z} \gg \kappa_{\rm r}$ in the 
halo\footnote{Note that in the disk, $\kappa_{\rm r}$, and in the halo, 
$D_{\rm z}$, are both {\em parallel} to the large scale regular magnetic field 
direction.}, 
with $D_{\rm z}$ and $\kappa_{\rm r}$
being the components of the diffusion tensor perpendicular and parallel to 
the Galactic plane.
Thus, in general, CRs spend a considerable amount of their transport time
in the halo, subject to {\em anisotropic} diffusion.

\section{A simple model for the Galactic CR source distribution}
\label{sources}
The inference on the spatial distribution of CR sources from {\em direct}
observations is plagued by a number of problems. From energy requirements for
the bulk of CRs below $10^{15}$ eV, it is known
that the only non-hypothetical Galactic candidates for the sources are 
SNRs and pulsars,  
with global energetic requirements favouring the former.  
SNRs can be best studied in the radio continuum and in soft X-rays, but as 
low surface brightness objects larger and older remnants are systematically
missed. Since samples are usually flux limited, the more distant objects 
will be lost as well.
Pulsars on the other hand should be present in the Galaxy in large 
numbers. However they
are only detectable if their narrow beams happen to cross the
line of sight of the observer or, in X-rays, as isolated old neutron stars; 
so far only four candidates are known from the ROSAT All-Sky Survey 
(Neuh\"auser \& Tr\"umper 1999). Therefore
selection effects will bias samples heavily in both cases.

Based on the observations of SNRs by Kodeira (1974) and pulsars by
Seiradakis (1976), Stecker \& Jones (1977) have given a simple radial
distribution of the form
\begin{equation}
Q_1(x) = q_0 x^a \exp(-b x) \,,
\label{sjd1}
\end{equation}
where, $Q_1(x)$ is a SNR surface density in $x$, with  
$x=r/R_{\odot}$, and $a$ and $b$ are matched to the observations. 
Here we adopt $a=1.2$ and $b=3.22$.
Since both SNRs and pulsars are associated
with Population~I objects, it is reassuring that $a$ and $b$ are very
similar for both classes of objects. 

\begin{figure}[thbp]
   \centering
   \mbox{%
   \includegraphics[width=\hsize,clip]{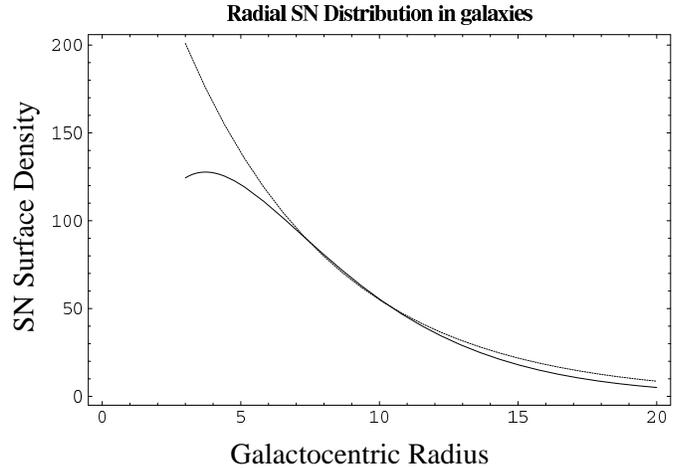}
}
      \caption[]{Surface density of SNRs and supernovae, respectively, 
      in galaxies as a function 
      of Galactocentric distance $R$. The solid curve, $Q(R)$, gives 
      a fit to the SNR distribution of the Milky Way and is given by 
      $Q(R) = 87.45 R^{1.2} 
      \exp[-0.322 R]$, whereas the dotted curve is a fit to the extragalactic 
      supernova distribution by Dragicevich et al. (1999), given by $S(R) = 350 \, 
      \exp[-0.185 R]$. Note the excellent agreement in the exponential decrease  
      of $Q(r)$ and $S(R)$ for $5 < R < 20$ kpc. %
      }
         \label{snd4}
   \end{figure}

To determine the proportionality constant $q_0$ for the distribution
given in Eq.~(\ref{sjd1}), we use the results obtained by Leahy \& Xinji (1989)
who used the catalogue of Li (1985) derived from radio observations.
Leahy \& Xinji (1989) considered shell-type SNRs and applied empirical
correction factors due to the incompleteness of the flux limited sample.
The spatial distribution thus obtained shows a peak at 4-6 kpc from the
Galactic center (see Fig.~\ref{snd4}), assuming an overall rotational symmetry. 
A more systematic study has recently been undertaken 
by Case \& Bhattacharya (1998), who have made a careful analysis of an 
enhanced Galactic SNR sample, using an improved $\Sigma-D$ relation. 
These authors 
find a peak in the distribution at around 5 kpc\footnote{The difference is 
due to the fact that with the older value of $R_{\odot} = 10$ kpc 
in Eq.~(\ref{sjd1}) the peak 
of the distribution appears at $r_{\rm max} = {a\over b} R_{\odot} 
\sim 4$ kpc} and a scale length of $\sim 7$ 
kpc, thus confirming the gross features of the older work. 

The number of SNRs in an annulus of width $dx$ is given by
\begin{eqnarray}
dN &=& Q_1(x) 2 \pi x dx \nonumber \\
&=& 2 \pi q_0 x^{a+1} \exp(-b x) dx \,,
\label{sjd2}
\end{eqnarray}
and thus the total number of SNRs in the disk of radius $\sim 18$~kpc  
is
\begin{eqnarray}
N &=& 2 \pi \int_{0}^{x_{\rm g}} q_0 x^{a+1} \exp(-b x) dx 
\label{nsn2} \\
&\approx& 2 \pi q_0 \int_{0}^{x_{\rm g}} x^2 \exp(-b x) dx \nonumber \\
&=&2 \pi q_0 \left[-\exp(-b x) \left({x^2 \over b} + {2 x \over b^2}
+ {2 \over b^3} \right)\right]_{0}^{x_{\rm g}} \nonumber \\
&\approx& 0.35 \, q_0 \,,
\label{nsn1}
\end{eqnarray}
for $x_{\rm g}=1.8$, $a\approx 1$ and $b=3.22$. 
Numerical integration of Eq.~(\ref{nsn2}) yields $N = 0.33 \, q_0$.

On the other hand, the total number of SNRs that should be observable at
present is their production rate $\theta$ times their average life
time $\tau_{\rm SN}$ before they merge with the hot intercloud medium and
lose their individual appearance, or observationally escape the 
detection limit. The rate at which SNe occur in the disk 
is $2.2\pm 1.3$ per century for randomly exploding stars (van den 
Bergh 1990, 1991) and 0.45 per century
for explosions occurring in OB associations (Evans et al. 1989), 
giving a total SN rate in the Milky Way disk of 
$\theta = 2.65$ per century. 
The total number of SNRs in the disk is then given by
\begin{equation}
N = \theta \, \tau_{\rm SN} \,;
\label{totsnr}
\end{equation}
 for $\tau_{\rm SN} = 1.83 \, 10^4$ years we reproduce  
the value given by Leahy \& Xinji (1989), who estimate
$N=485 \pm 60$ for Galactic SNRs with a limiting surface brightness of
$\Sigma \ge 3 \times 10^{-22} \, {\rm W} \, {\rm m}^{-2} \, {\rm Hz}^{-1} \,
{\rm sr}^{-1}$.
 From Eq.~(\ref{nsn1}) we obtain: $q_0 = 1386\pm 171$.

Dragicevich et al. (1999) have analyzed also the radial distribution of 
SNe in a sample of 218 
{\em external galaxies} of different Hubble types, corrected for inclination 
angle. The data were sorted into radial bins and the numbers converted into 
SNe surface densities. 
They showed that the radial SN surface distribution can be well fitted by an 
exponential radial decrease of the form
\begin{equation}
S(R) = \sigma_{\rm c} \, {\rm e}^{-R/R_{\rm sc}} \,, 
\end{equation} 
\label{sjd3}
with $\sigma_{\rm c} = 350$. Taking a sub-sample of 36 Sb-Sbc galaxies, 
they were able to derive a value of $R_{\rm sc} = 5.4$ kpc for the exponential 
scale length of the 
Milky Way as a representative Sb-Sbc galaxy. The number of SNe derived is 
consistent with 
historical records of Galactic SNe within 4 kpc from the Sun. 
In Fig.~\ref{snd4} it is shown that both the exponential decrease and, 
interestingly, but rather fortuitously, also the 
absolute numbers in the SN surface density 
agree fairly well for $R > 5$ kpc with the earlier 
derived relation~(\ref{sjd1}) for the SNR surface density. 
  
Therefore the value calculated for the proportionality 
constant, $q_0$, provides a reliable estimate for the number of Galactic SNRs 
in a circular ring at a distance $R$. 
   
It should be emphasized that the SNR lifetime, $\tau_{\rm SN}$, does not 
enter the analysis of ~Sect.~\ref{radvar}, (see Eq.~(\ref{pcx1})), since 
it is only the CR production rate and hence the SNR rate per unit volume that 
determines the truly diffuse $\gamma$-ray emissivity. 

\section{Flattening of the radial $\gamma$-ray emissivity gradient}
\label{flagra}

In the following we will argue that the CR propagation picture 
needs to be changed for two physical reasons.
Firstly, as we have already pointed out, diffusive transport in the disk 
and in the halo is not the same due to a different magnetic field geometry 
and to different wave excitation and damping mechanisms. Thus {\em anisotropic}
diffusion seems to be rather the rule than the exception. Secondly, advection
of CRs is the dominant transport mode above a certain transition boundary 
located at
($R_{\rm C}, z_{\rm C}$), with $z_{\rm C} = z_{\rm C}(r)$ varying with Galactic radius.

More specifically, we
consider SNRs as the primary sources for the Galactic CRs. Each remnant
results from an energy deposition of $E_{\rm SN} \simeq 10^{51} \, {\rm erg}$, 
which is converted into relativistic particles with a certain efficiency
$\nu \sim 0.1 - 0.5$ (Drury et al. 1989; Berezkho \& V\"olk 2000); 
the individual value of $\nu$ is poorly known from theory due to 
uncertainties in the injection process and depends e.g., on the value and 
orientation of the circumstellar magnetic field.
SNe exploding inside a superbubble, i.e.\ a hot tenuous medium, initially  
generate low 
Mach number shocks, which are less efficient in accelerating particles. 
After a time of the order of the sound crossing time, however, the shock 
impinges on the much colder and denser surrounding shell and becomes progressively 
stronger thereby accelerating particles more efficiently; this should 
to lowest 
order compensate for the initially decreased efficiency in the hot ambient 
medium. We would expect that 
due to the continuous energy input by successive SN explosions 
also a long lived shock would be able to accelerate particles to energies 
in excess of $10^{14}$ eV, albeit adiabatic energy losses  
would become more and more severe with time. Since the diffusion coefficient of CRs 
increases with energy, advective transport of particles significantly above 
1 GeV will be the dominant mode of transport only at larger distances from 
the Galactic plane. Therefore these particles will quickly fill an 
extended halo and not generate many $\gamma$-rays in the disk via $\pi_0$-decay. 

Bykov \& Fleishman (1992) have argued that successive explosions inside 
a bubble can generate strong turbulence, which should transform a 
significant amount of the total free energy to cosmic rays. However, 
at the same time the injection rate at the shocks may be reduced as a 
result of shock modification due to previously generated CR particles.    
With the details of the acceleration mechanism in superbubbles being 
still debatable, we believe that 
to lowest order there is no difference in the overall energy transfer from 
thermal plasma to CRs, if a SN explodes inside a superbubble or just forms a 
single remnant. Thus the energy production rate of CRs should be roughly 
proportional to the number of SN explosions, regardless whether they 
occur in the general ISM or inside a superbubble. 
The numerical difference in the derived CR energy density (and CR pressure) 
(cf.\ Eqs.~(\ref{enden1}) and (\ref{snrate})), however, between treating particle 
acceleration in superbubbles as equal to single remnants, and disregarding 
acceleration in superbubbles altogether, is small. 
According to the previous Sect. it amounts to a factor 
$\theta_{\rm tot}/\theta_{\rm SNR} = 2.65/2.2 \approx 1.2$, and is therefore well 
below the uncertainty in the acceleration efficiency $\nu$.
In the following, we tend to be conservative and retain a low value of 
$\nu = 0.1$. 

Using the results from Sect.~\ref{sources} we can estimate the local 
(in radius) number of
SNRs within a circular ring of the Galaxy with a width of, say, 2 kpc.
Relating the SN rate directly to the number of observable SNRs by 
Eq.~(\ref{totsnr}) and writing (cf.\ Eq.~(\ref{sjd2}))
\begin{equation}
Q_1(x)={{\Delta N(x)}\over{2\pi x \Delta x}}\,,
\label{nosodi1}
\end{equation}
we can now define the SNR rate per unit radial confinement volume, 
$\eta(x)$, by
\begin{eqnarray}
\eta(x) &=& {\Delta N(x)\over \tau_{\rm SN}\, \Delta V} \nonumber \\
        &=& {\theta \, Q_1(x) \over N z_{\rm C} R_{\odot}^2} \,.
\label{snrate}
\end{eqnarray}

The local CR energy density, $\epsilon_{\rm C}$ in the disk at $z \approx 
0$, can then be estimated to be
\begin{equation}
\epsilon_{\rm C}(x) = {\gamma_{\rm C} \over \gamma_{\rm C}-1} P_{\rm C}(x)
 = {1 \over 2} \, \eta(x) \, E_{\rm SN} \, \nu \, \tau_{\rm esc} \,,
\label{enden1}
\end{equation}
with $\Delta V(x) = 2 \pi x \Delta x z_{\rm C} R_{\odot}^2$, and $\tau_{\rm esc}$ 
being the local diffusive CR storage volume and time, respectively; the 
factor $1 \over 2$ accounts for one half-space, e.g.\ $z>0$. The CR pressure 
in the disk is denoted by $P_{\rm C}(x)$ and $\gamma_{\rm C} \approx 4/3$ is 
the specific heat ratio for the CRs.
Working in the framework of an average transition boundary, $z_{\rm C}$, which 
divides space below and above in purely diffusion and advection regions, 
respectively, the probability for particles at $z > z_{\rm C}$ to return and hence 
contribute to the CR pressure in the disk becomes exponentially small. We have 
therefore used $z=z_{\rm C}$ as the vertical extension for the CR storage volume. 
As we can see from Eq.~(\ref{snrate}), $\eta(x)$, depends on the 
local and total number of sources and the global SN rate, but not on the 
SNR life time.
The vertical extent of the diffusion-advection boundary $z_{\rm C}$ is determined
by the balance of diffusive and advective CR flux at this vertical distance, 
i.e.\ to lowest order by
\begin{equation}
z_{\rm C} \sim {D \over V}
= {D \over u_0 + V_{{\rm A} 0}} \,,
\label{reflev}
\end{equation}
and the local CR escape time is
\begin{equation}
\tau_{\rm esc}(x) \sim {z_{\rm C}^2(x) \over D} \sim {D \over 
(u_0(x) + V_{{\rm A} 0}(x))^2} \,
\label{locesct}
\end{equation}
Thus the radial CR pressure variation is given by
\begin{eqnarray}
P_{\rm C}(x) &=& {\gamma_{\rm C}-1 \over 2 \gamma_{\rm C}} 
{\eta(x) \nu \, E_{\rm SN} \, z_{\rm C}  \over (u_0(x) + V_{{\rm A} 0}(x))} \,, 
\nonumber \\
&=& {\gamma_{\rm C}-1 \over 2 \gamma_{\rm C}} 
{\theta \, Q_1(x) \nu \, E_{\rm SN} \over N 
\, R_{\odot}^2 \, (u_0(x) + V_{{\rm A} 0}(x))} \,,
\label{pcx1}
\end{eqnarray}
It is easy to show that Eq.~(\ref{pcx1}) follows directly from the kinetic 
equation for CRs, as can be seen from Eq.~(\ref{pcsol3}).

The diffuse $\gamma$-ray intensity resulting from $\pi^0$-decay photons is
\begin{equation}
I_{\gamma} \propto \int_0^{l} \, n_{\rm C} \, n_{\rm H} dl
\sim \langle n_{\rm C} \rangle \, \langle n_{\rm H} \rangle \, l \,,
\label{iga1}
\end{equation}
with $\langle n_{\rm C} \rangle$ and $\langle n_{\rm H} \rangle$ being the 
line-of-sight averaged CR and hydrogen number 
density, respectively, and $l=\sqrt{z_{\rm C}^2+(r-R_{\odot})^2}$ is the line 
of sight. 
 From Eq.~(\ref{pcx1}) we have
\begin{equation}
n_{\rm C} \propto P_{\rm C}(x) \propto {Q_1(x) \over u_0 + V_{{\rm A} 0}} \,,
\label{pr_rat}
\end{equation}
disregarding the changes in the CR distribution function 
due to energy-dependent diffusion.      
$n_{\rm H}$ is an observationally fixed quantity, for which we 
have to use its line of sight averaged value (see Eq.~(\ref{iga1})) 
if we are interested in 
quantitative calculations of the $\gamma$-ray intensity; here we are 
primarily concerned with the radial variations of $I_\gamma$.

In order to obtain $n_{\rm C}\approx const$ for a weak $\gamma$-ray gradient,   
we should assume that
\begin{equation}
(u_0 + V_{{\rm A} 0}) \propto Q_1(x) \,,
\label{invel1}
\end{equation}
if $V \sim {\rm const}$;  if $V = V_0 z$, we have 
$z_{\rm C} \sim \sqrt{D/V_0}$ and therefore should require 
\begin{equation}
(u_0 + V_{{\rm A} 0}) \propto Q_1^2(x) \,.
\label{invel2}
\end{equation}

Combination of Eqs.~(\ref{locesct}) and (\ref{invel1}) leads to
\begin{equation}
\tau_{\rm esc}
\propto {D \over Q_1^2(x)} \,.
\end{equation}
All other things being equal, a locally enhanced CR production rate should
therefore lead to a reduced CR escape time and diffuse $\gamma$-ray
intensity.

\section{Radially Dependent Cosmic Ray Driven Outflows}
\label{radvar}
As we have deduced in the last Sect., the CR pressure $P_{\rm C}(x)$ in 
the disk is a 
radially dependent quantity, and therefore we expect the outflow 
velocity, $u(x)$, and the mass loss rate, $\dot m(x)$, to be also radially 
dependent. In an earlier paper (Breitschwerdt et al. 1991) we have 
shown that such a behaviour already exists as a consequence of the radial 
dependence of the gravitational potential. The net result was a monotonic 
decrease (increase) of terminal velocity (mass loss rate) with increasing 
Galactic radius for a radially constant mass density, $\rho_0$. 
Now we have superposed the radial variation of the CR source 
density $Q_1(x)$ and investigate in the following how this changes 
the outflow. 
 
However, Eq.~(\ref{pcx1}) is an {\em implicit} equation, since $P_{\rm 
C}(x)$ depends 
on $u_0(x) + V_{{\rm A} 0}(x)$, which in turn depends on the energy 
density available in CRs, i.e.\ $\epsilon_{\rm C}(x)$ and thus $P_{\rm 
C}(x)$ itself, to drive the outflow. 
To that end we have performed {\em self-consistent} galactic wind 
calculations of the fully nonlinear equations, in which for a given 
gravitational potential of the Milky Way, 
and a relativistic CR gas ($\gamma_{\rm C} = 4/3$),   
together with a spatially averaged mass density $\varrho(x) \sim \varrho_0 = 
1.67 \times 10^{-27} \, {\rm g}\, {\rm cm}^{-3}$, an average thermal 
pressure $P_{\rm g}(x) \sim P_{\rm g0} = 2.76 \times 10^{-13} \, {\rm dyne}\, 
{\rm cm}^{-2}$, an averaged halo magnetic field $B(x) \sim B_0 = 1 \mu{\rm G}$, 
and a small average level of wave amplitude $\delta B(x)/B(x) \sim 
\delta B_0/B_0 = 0.1$, the advection-diffusion boundary $z_{\rm C}(x)$, 
$P_{\rm C}(x)$ and $u_0(x) + V_{{\rm A} 0}(x)$ are calculated 
self-consistently, using $q_0 = 1492$, the value derived from the numerical 
integration of Eq.~(\ref{nsn2}) (see Sect.~\ref{sources}).  
The form of the potential (including, disk, bulge and dark matter halo) and 
the opening of the flux tube due to geometrical divergence are the same as 
used by Breitschwerdt et al. (1991). 
   \begin{figure}[thbp]
   \centering
   \mbox{%
   \includegraphics[width=0.9\hsize,angle=-90,clip]{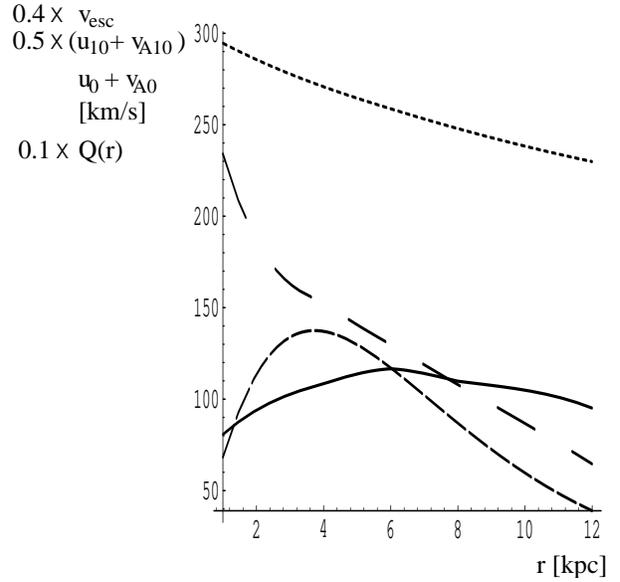}
}
      \caption[]{Escape speed, $v_{\rm esc} = \sqrt{-2 \Phi_0}$, 
      of the Galaxy (dotted curve), 
      outflow velocity, $u_{10}+v_{{\rm A} 10}$, at a vertical distance of 
      $z=10$ kpc (long dashed curve), 
      and outflow velocity, $u_{0}+v_{{\rm A} 0}$, at reference level, 
      $z_{\rm C}$ (solid curve), 
      respectively, as seen in a fixed frame of reference, and the 
      supernova distribution function (short dashed curve), 
      $Q(r)$, (see Eqs.~(\ref{sjd1}) and (\ref{nsn1}), as a function of 
      galactocentric radius $r$. 
      Curves have been scaled down by the factors indicated on the 
      vertical axis to fit into one diagram.                
      For boundary conditions of self-consistent galactic wind calculations 
      see text. 
}
         \label{snd1}
   \end{figure}
   \begin{figure}[thbp]
   \centering
   \mbox{%
   \includegraphics[width=\hsize,angle=-90,clip]{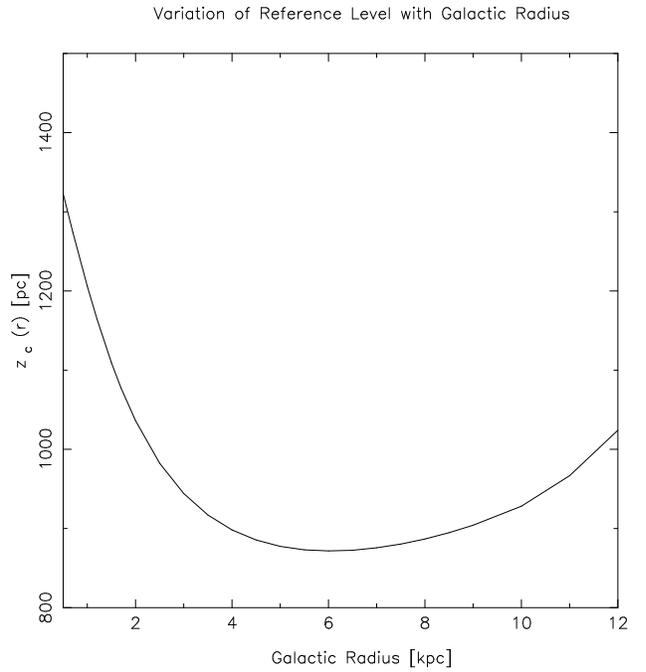}
}   
      \caption[]{Dependence of the diffusion-advection boundary, $z_{\rm C}(r)$, 
      on Galactocentric radius $r$. The location of $z_{\rm C}$ at any 
      radius is defined by the balance of diffusive and advective CR fluxes. 
      The curve shown here has been obtained from self-consistent galactic 
      wind calculations (see text). }
         \label{snd3}
   \end{figure}
We have used a radial grid for our calculations in the Galactic 
radius interval from $0.5 - 12$ kpc, with spacings of $0.5$ kpc. 
The net radial dependence of the outflow velocity we obtained 
from our numerical calculations is shown  
in Fig.~\ref{snd1}, and its $r$- and $z$-dependence in Fig.~\ref{snd2}. 
The outflow velocity $u_0 + V_{{\rm A} 0}$ in Fig.~\ref{snd1} (solid line) 
for a fixed Galactic frame of reference at 
the radially varying transition boundary $z_{\rm C}(r)$, shows a similar convex 
shape as the source distribution $Q_1(x)$ (short dashed curve), although much 
less pronounced. 
This is due to the fact, that the decrease of the gravitational potential, 
as can be seen from the escape speed, $v_{\rm esc} = \sqrt{-2 \Phi_0}$, 
with $\Phi_0 = \Phi_{\rm b} + \Phi_{\rm d} + \Phi_{\rm h}$ for bulge, disk and halo 
contributions, respectively, (dotted curve in Fig.~\ref{snd1}),  
keeps the outflow velocity from dropping off too rapidly with increasing 
Galactocentric distance. Therefore, even a uniform SNR distribution would 
enforce a radially varying outflow velocity, as has been shown in Fig.~7 of 
Breitschwerdt et al. (1991). The radial variation of the outflow velocity 
at a distance of $z=10$ kpc from the disk, $u_{10} + V_{{\rm A} 10}$, 
is shown by the long dashed curve, which both shows the transition from 
the flow velocity at reference level $z_{\rm C}$ to the escape speed, and also 
a roughly $1/r^2$-dependence, which we will use in our analytical treatment 
later (cf.~Eq.~(\ref{velvar1})).

It can be seen from comparison of $z_{\rm C}(r)$ and $u_0(r) + V_{{\rm A} 0}(r)$, 
as obtained from the fully nonlinear calculations, that the simple 
ansatz of Eq.~(\ref{reflev}) is indeed fulfilled. The functional dependence 
of $z_{\rm C}(r)$ with radius is straightforward to understand. Close to the 
Galactic centre, the gravitational pull is strongest, as can be seen from 
$v_{\rm esc}$ in Fig.~\ref{snd1}; since we chose all other quantities 
being the same across the disk (constant density, thermal pressure, 
magnetic field strength), the outflow velocity, and hence mass loss rate, 
are smallest here. Eq.~(\ref{reflev}) then tells us, that $z_{\rm C}(r)$ must 
be largest. At the outer parts of the Galaxy, the gravitational field 
becomes weaker, but also the source distribution, and hence the CR pressure, 
decrease exponentially, and so the outflow velocity (cf.~Fig.~\ref{snd1}) 
also decreases, and $z_{\rm C}(r)$ must increase again (see Fig.~\ref{snd3}). 
It is noteworthy that 
the maximum of $u_0(r) + V_{{\rm A} 0(r)}$ and the minimum of $z_{\rm C}(r)$ at 
$r \sim 6$ kpc {\em do not coincide} with the maximum of $Q_1(r)$ and 
$P_{\rm C}(r)$, respectively, at $r \sim 3$ kpc. This must be a consequence 
of the interplay between the gravitational field 
and the source distribution in the fully nonlinear equations (cf. 
Breitschwerdt et al. 1991). 

Finally we mention that the choice of constant boundary conditions across 
the Galactic disk is a conservative one. In reality we should also enhance 
the thermal temperature and pressure in regions of higher supernova activity. 
The net effect would be a more pronounced peak in outflow velocity and 
a deeper minimum in $z_{\rm C}(r)$, respectively, and therefore an even ``better''
quantitative proportionality between $u_0(r) + V_{{\rm A} 0(r)}$ and 
$Q_1(x)$ according to Eq.~(\ref{invel1}).

\section{Models of advective and anisotropic diffusive CR transport}
\label{crtrans}
In the following we discuss several advection models
(some including also anisotropic spatial diffusion), in which 
we examine in detail the ideas presented in Sect.~\ref{flagra}. We show that 
for a given radial CR source distribution function $f_1(r) \propto Q_1(r)$, 
we can find a function $f_2(r) \propto V(r)$, for which the effect of the 
radially varying Galactic wind velocity leads to an almost uniform CR 
distribution in
the disk. We refer to relations $f_2(r) = F(f_1(r))$ as ``compensation 
equations''. It is thus possible to explain the observational data 
by pure CR propagation effects. 

To fix ideas, we start out with a functional relation for the Galactic CR 
source distribution of the form
\begin{equation}
Q(z, r, E) = K(E) \, f_1(r) \,\delta (z)\,,
\label{q}
\end{equation}
neglecting the thickness of the Galactic disk; here $\delta (z)$ 
denotes the Dirac delta function. Then for a given source
distribution function, $f_1(r)$, we try to find a wind structure of the form 
\begin{equation}
V(r,z)=V(z) \, f_2(r)\,,
\label{v}
\end{equation}
which makes the CR distribution almost uniform. In the following 
we describe models of increasing complexity with respect to geometry 
and outflow velocity.   

\subsection{One-dimensional model with a constant galactic wind velocity}
We start from the simplest one-dimensional CR transport model, which is an 
extreme case of anisotropic diffusion, since only propagation in the 
$z$-direction is allowed. In this case the galactocentric radius $r$ is 
simply a parameter in the model.

The one-dimensional diffusion-advection equation for CR nucleons can be 
written as
\begin{equation}
{\partial\over{\partial z}}\left(V(r,z) \, n - D{{\partial n}\over{\partial z}} 
\right)-{1\over 3}{{dV}\over{dz}}{\partial\over{\partial E}}(E \, n) = 
Q(E,r,z)\,,
\label{one-dim}
\end{equation}
where for simplicity we take the advection velocity to be constant 
and directed away from the Galactic plane
\begin{equation}
V(r,z)=\pm V_0 \, f_2(r) \, \Theta(\pm(z-z_{\rm C}))\,.
\end{equation}
Here $\Theta(z-z_{\rm C})$ is the Heavyside step function, and the sign "$+$" 
corresponds to the velocity above, and the sign "$-$" to the one below the 
Galactic plane.  

The boundary conditions for CRs are determined either by free escape 
into intergalactic space, if the density of electromagnetic fluctuations 
generated by the CR flux decreases fast enough (see Dogiel et al. 1994), or 
by CR advection in a galactic wind to infinity (Breitschwerdt et 
al. 1991; Ptuskin et al. 1997), if the level of fluctuations is high 
enough. 
Which of these cases is relevant for the Galaxy, is the subject of a 
separate investigation. Fortunately, CR spectra and densities in 
the disk are independent of the boundary conditions {\em far away} from the 
Galactic plane, if there is an outer region of advective transport. 
In the case we discuss here, it is assumed that the CR propagation region can 
be formally divided into a diffusion 
halo wrapping around the Galactic plane and an adjacent advection region, 
reaching out to intergalactic space.  
These two regions are separated by a boundary surface, $z_{\rm C}$, at which both the 
CR density and the flux have to be continuous.
 
The sources are concentrated in the disk 
and are supposed to emit a power-law spectrum of particles
\begin{equation}
Q(E,r,z)=z_{\rm d} \, K \, f_1(r) \, E^{-\gamma_0} \, \delta(z)\,,
\end{equation}
where $z_{\rm d}$ is the thickness of the Galactic disk and the coefficient $K$ 
determines the CR production by SNRs in the disk. 

The location of the transition boundary 
for a constant diffusion coefficient $D$ and advection velocity $V_0$ is 
determined in this approximation by
\begin{equation}
z_{\rm C} = {D\over V_0} \,.
\end{equation}
Then for $z<z_{\rm C}$ the transport equation reads
\begin{equation}
-D{{d^2n_{\rm d}}\over{dz^2}}=z_{\rm d} \, K \, f_1(r) \, E^{-\gamma_0} \, \delta(z) \,,
\label{deq1}
\end{equation}
and for  $z\ge z_{\rm C}$
\begin{equation}
{d\over{dz}}(n_{\rm C} V_0 f_2(r))+{1\over 3}V_0 f_2(r)\delta(z-z_{\rm C})(\gamma_0-1) \, 
n_{\rm C}=0 \,.
\label{ceq1}
\end{equation}
Since adiabatic energy losses do not change the particle spectral index, 
$n\propto E^{-\gamma_0}$.
The solution of Eq.~(\ref{deq1}) is 
\begin{equation}
n_{\rm d}=C_1+C_2 z
\end{equation}
and the CR density is constant in the advection region, 
\begin{equation}
n_{\rm C}=C_0 \,.
\end{equation}
The constants can be determined from the boundary conditions at $z=0$
\begin{equation}
-D{{d n_{\rm d}}\over{dz}}=z_{\rm d} \, K f_1(r) E^{-\gamma_0}\,,
\end{equation}
and at $z=z_{\rm C}$ by continuity of the diffusive and advective CR fluxes 
\begin{equation}
D{{d n_{\rm d}}\over{dz}} = V_0 f_2n_{\rm C}+{{V_0f_2}\over 3} (\gamma_0-1) n_{\rm C} \,.
\end{equation}
Since at $z=z_{\rm C}$ also $n_{\rm d} = n_{\rm C}$, we obtain   

\begin{equation}
n_{\rm C}={{3z_{\rm d} K E^{-\gamma_0}}\over{V_0 (\gamma_0+2)}} {f_1(r) \over f_2(r)} \,,
\end{equation}
and therefore 
\begin{equation}
n_{\rm d}=z_{\rm d} K f_1(r) E^{-\gamma_0}\left({3 \over{V_0 f_2(r) (\gamma_0+2)}} - 
{z-z_{\rm C}\over D} 
\right)\,.
\label{ndeq1}
\end{equation}
We see that at $z=0$ 
\begin{equation}
n_{\rm d}={{z_{\rm d} K f_1(r) E^{-\gamma_0}}\over{f_2
V_0}}{{(\gamma_0+5)}\over{
(\gamma_0+2)}}\,,
\end{equation}
and the ``compensation'' equation must have the form 
\begin{equation}
f_1(r) = f_2(r) \,,
\end{equation}
for the CR particle density to be uniform, in other words, the radial 
dependence of the outflow velocity must be the same as that of the 
source distribution. 

 From Eq.~(\ref{ndeq1}) we can also estimate the total pressure, $P_{C}$, 
of CRs in the disk, which is
\begin{eqnarray}
P_{C} &=& {\int\limits_{E_0}^\infty} n_{\rm d}(z=0,E)E\cdot dE \\
&=& {{z_{\rm d} K f_1 E_0^{-(\gamma_0-2)}}\over{(\gamma_0-2) f_2 V_0}}
{{(\gamma_0+5)}\over (\gamma_0+2)} \,.
\label{pcsol1}
\end{eqnarray}
The physicals units of $K$ are $[K] = {\rm erg}^{\gamma_0-1}/{\rm s}$, 
and $K E_0^{-(\gamma_0-2)}$ equals the CR power in the Galactic disk, 
hence (neglecting constants of order unity)
\begin{equation}
K E_0^{-(\gamma_0-2)} \simeq \nu E_{\rm SN} {\theta \over N} \,.
\end{equation}
On the other hand the normalized source distribution (cf.\ Eq.~(\ref{nosodi1}))
is equivalent to 
\begin{equation}
{Q_1(x) \over  R_{\odot}^2} = {\Delta N \over \Delta A_{\rm d}} = {f_1 \Delta V_{\rm d} 
\over \Delta A_{\rm d}} = f_1 z_{\rm d} \,,
\end{equation}
where $A_{\rm d}$ and $V_{\rm d}$ are the area and volume of the Galactic disk, 
respectively, 
and identifying $V_0$ with $u_0+V_{{\rm A} 0}$, Eq.~(\ref{pcsol1}) becomes
\begin{eqnarray}
P_{\rm C} &=& {z_{\rm d} f_1} {K E_0^{-(\gamma_0-2)} \over{(\gamma_0-2) f_2 V_0}}
{{(\gamma_0+5)}\over {(\gamma_0+2)}} \nonumber\\
&\simeq&
{Q_1(x) \over  R_{\odot}^2} \cdot \nu E_{\rm SN} {\theta \over N
(u_0+V_{{\rm A}0})} \,,
\label{pcsol3}
\end{eqnarray}
which corresponds to Eq.~(\ref{pcx1}), thus verifying the 
validity of the intuitively derived relation between CR pressure, source 
distribution and Galactic SN rate in a Galactic wind flow.

\subsection{One-dimensional model with $z$-dependence of the wind}

Based on the conclusions of the previous Sections, we can generalize the
one-dimensional solution obtained by Bloemen
et al. (1993), taking  into account  radial variations of the sources and 
the wind velocity. Let us suppose that the advection velocity has the form 
\begin{equation}
V(z,r) =3 \tilde V_0 z f_2(r) \,; 
\end{equation}
the solution of Eq.~(\ref{one-dim}) is then given by
\begin{equation}
n(z=0,r)\simeq A{{K f_1(r)} \over {\sqrt{D \tilde V_0 f_2(r)}}}E^{-\gamma} \,,
\end{equation}
where $A$ is a constant; note that $\tilde V_0$ is a typical time 
constant (e-folding time) of the flow.

We find that a vanishing radial dependence of the CR distribution, $n=const$, 
can be obtained if 
\begin{equation}
f_1^2(r) = f_2(r)\,.
\end{equation}

In these simple models the size of the halo can be derived only locally,  
because global or large-scale characteristics, like e.g.\ the CR distribution 
along the Galactic plane, are independent of the CR propagation itself.

These simple analytical solutions of the one-dimensional transport 
equation show that the main effect, which formally leads to the 
"compensation", is a curved transition boundary between diffusion and 
advection regions, i.e.\
\begin{equation}
z_{\rm C}(r)\sim {D \over V(r)} \,;
\end{equation}
note that such a behaviour is indeed apparent from Figs.~\ref{snd1} and 
\ref{snd3}.  
The main point here is that the compensation effect is a consequence of the 
$r$-dependence of the CR life-time, $z_{\rm C}^2/D\sim D/V^2(r)$.

In the framework of the pure one-dimensional model it is unimportant 
how far from the Galactic plane the
boundary $z_{\rm C}(r)$ is, but as we shall see, this figure will be essential 
for the three-dimensional case.

\subsection{Three-dimensional axisymmetric model with radial
velocity dependence}
To demonstrate the effect of radial changes of the boundary $z_{\rm C}(r)$ between
regions of diffusive and advective propagation of CRs, we
investigate the diffusion-advection equation with a given radial dependence 
of the wind velocity and the source density in a more realistic cylindrical 
geometry (axial symmetry, i.e.\ $\partial/\partial \varphi \equiv 0$) with the
velocity varying as
\begin{equation}
V(r)={V_0\over r} = V_0 f_2(r)\,,
\label{veldep1}
\end{equation}
and the source distribution as
\begin{equation}
Q(r,z)={Q_0\over {r^\alpha}} = Q_0 f_1(r)\,.
\end{equation}
Then the equation for the CR distribution function we have to solve reads 
\begin{equation}
{V_0\over r}{{\partial n}\over{\partial z}}= D_{\rm z}{{\partial ^2} \over
{\partial z^2}} n + \kappa_{\rm r} {1\over r}{\partial\over{\partial r}}
\left(r {\partial n \over \partial r}\right) + {Q_0\over 
{r^{\alpha}}}\delta (z) \,,
\label{diff-conv2}
\end{equation}
where we have stressed the anisotropic diffusion by two different constant
diffusion coefficients in the radial and perpendicular direction, 
$\kappa_{\rm r}$ and $D_{\rm z}$, respectively. Note that $V_0$ has the same dimensions
as the diffusion coefficients, and that $n(r,z)$ is the particle distribution 
function in phase space. The boundary conditions at $z=0$ are implicitly 
included in Eq.~(\ref{diff-conv2}).
Here we treat a special case of radial variation of the boundary $z_{\rm C}$:
\begin{equation}
z_{\rm C}(r)\sim r{D_{\rm z}\over V_0} \,,
\label{trans-bound1}
\end{equation}
corresponding to the velocity dependence given in Eq.~(\ref{veldep1}).
Comparison with Figs.~\ref{snd1} and \ref{snd3} shows that such a behaviour 
may indeed represent the outer Galaxy, where the outflow velocity decreases 
with increasing radius and the distance of the boundary $z_{\rm C}$ from the 
Galactic disk increases accordingly. 

Eq.~(\ref{diff-conv2}) can be solved analytically 
if we restrict ourselves to self-similar solutions, in which 
the distribution function does not vary with $r$ and $z$ independently.
The price we have to pay is that there is no unique solution that 
covers both $0\leq r < \infty$ and $0\leq z < \infty$. Instead 
a similarity solution is found for $z \to 0$ and $r>0$, and 
one for $z>0$, which have to match in overlapping 
regions. We start out with the latter one, applying a   
transformation of independent variables in the form 
\begin{equation}
\xi= \sqrt{D_{\rm z}\over \kappa_{\rm r}}{r\over z} \,,
\label{z_xi}
\end{equation}
and introducing the function
\begin{equation}
\phi(\xi)=z^{-\mu} n(z,r) \,.
\label{phitraf1}
\end{equation}
Equation~(\ref{diff-conv2}) in terms of $\phi$ and $\xi$ then reads:
\begin{eqnarray}
\lefteqn{\xi(1+\xi^2){{\partial ^2 \phi}\over {\partial \xi ^2}}+(b_1 +
b_2\xi +b_3 \xi^2){{\partial \phi}\over{\partial \xi}}
 + }\nonumber\\
& &+(b_4 +b_5\xi)\phi - {Q_0 \over \kappa_{\rm r}}
{{r^{2-\alpha}}\over {z^{1+\mu}}}\delta(\xi-\infty)=0 \,,
\label{diff-conv3}
\end{eqnarray}
where the coefficients are given by
\begin{eqnarray}
\lefteqn{~~~b_1=1, ~~~~~b_2={V_0\over{\sqrt{\kappa_{\rm r} D_{\rm z}}}}, ~~~~~b_3=-2(\mu-1)}
\nonumber\\
&&b_4=-{{V_0 \mu}\over{\sqrt{\kappa_{\rm r} D_{\rm z}}}},~~~~~~~b_5=\mu(\mu-1)
\end{eqnarray}
This equation can be cast into a self-similar form for two sets of the
parameters $\alpha$ and $\mu$: $\alpha = 2$, $\mu= -1$, and $\alpha=
1$, $\mu = 0$.

We can derive an analytical solution of this equation in the form (see 
Bakhareva \& Smirnova 1980)
\begin{equation}
\phi(\xi)= K (1+\xi^2)^{C}\exp(A \arctan \xi) \,,
\end{equation}
where $K$ is a constant determined by the CR source power.

The unknown coefficients $\mu$, $C$ and $A$ are determined from the system
of algebraic equations which one can obtain from Eq.~(\ref{diff-conv3})
by equating the
coefficients for the different powers of $\xi$ to zero. The resulting system
of algebraic equations is given by 
\begin{eqnarray}
A b_1+b_4 &=& 0 
\label{algeq1} \\
2C (1+b_1)+A (A+b_2)+b_5 &=& 0 
\label{algeq2}\\
2C (2A+b_2)-A (2-b_3) + b_4 &=& 0 
\label{algeq3}\\
2C (2C-1+b_3)+ b_5 &=& 0 
\label{algeq4}\,.
\end{eqnarray}
We note that this system for the unknowns $A$, $C$ and $\mu$ is overdetermined;
therefore we have to check for consistency.

 From the system of equations (\ref{algeq1})-(\ref{algeq4}) it can be 
deduced that
\begin{eqnarray}
A &=& {V_0 \mu \over{\sqrt{\kappa_{\rm r} D_{\rm z}}}} \,, \\
C &=& {\mu \over 2} \,, \\
\mu_1 &=& 0 \,, \\
\mu_2 &=& -1 \,.
\end{eqnarray}
It can be verified that these solutions are all consistent.
However, $\mu=0$ implies $A=0$ and $C=0$, and hence does not yield any
physical solution.
Taking $\mu = -1$ gives $C=-{1\over 2}$. Therefore the solution is 
given by
\begin{equation}
\phi=K(1+\xi^2)^{-1/2}\exp(A \arctan \xi)\,,
\end{equation}
where the constant $K$ is determined from the conservation of particle
flux at $\xi=0$. 

The limit $z \to \infty$ corresponds to $\xi \to 0$ so that
\begin{eqnarray}
\lim_{\xi \to 0} \phi(\xi) &=& \lim_{\xi \to 0} K(1+\xi^2)^{-1/2} 
\exp(A \arctan \xi) \nonumber \\
&=& K = {Q_0 \over V_0} \,,
\end{eqnarray}
because at large distances from the disk ($z \to \infty$), 
the particle transport is completely determined 
by advective transport, and the number of particles ejected by the sources 
is conserved.  

The full solution of Eq.~(\ref{diff-conv2}) reads
\begin{eqnarray}
n(r,z) &=&  {Q_0
\over V_0 \left(z^2+{D_{\rm z}\over\kappa_{\rm r}} r^2 \right)^{1/2}}
\times \\
& &\exp\left\{-{V_0\over{\sqrt{\kappa_{\rm r} D_{\rm z}}}}\left(\arctan
\left[\sqrt{D_{\rm z} \over\kappa_{\rm r}}{r\over z}\right]\right)\right\} \,.
\label{diff-conv4}
\end{eqnarray}
 From the solution Eq.~(\ref{diff-conv4}) we can obtain the particle density 
behaviour at large $r$ and $z$, which are:
\begin{itemize}
\item[(i)] for $z\to\infty$:\\
\begin{equation}
n\propto {1\over z}\exp\left(-{V_0\over \kappa_{\rm r}}{r\over z}\right) \,,
\label{z_infty}
\end{equation}

\item[(ii)] and for $r\to\infty$:\\  
\begin{equation}
n\propto {1\over r}\exp\left(-{V_0\over D_{\rm z}}{z\over r}\right) \,.
\label{r_infty}
\end{equation}
\end{itemize}

 From Eq.~(\ref{diff-conv2}) it is clear that we are unable to 
describe the particle distribution near $z=0$ with the 
self-similar variable (Eq.~(\ref{z_xi})) and the function from 
Eq.~(\ref{phitraf1}), 
since it is impossible to satisfy the boundary conditions at $\xi=\infty$.

In order to study the radial behaviour of the distribution function near 
$z=0$ a different self-similar ansatz is used, which can be extended down 
to the spatial region near the sources. 
The similarity variable has the form   
\begin{equation}
\hat\xi= \sqrt{\kappa_{\rm r}\over D_{\rm z}}{z\over r} \,,
\label{r_xi}
\end{equation}
and the transformed distribution function reads
\begin{equation}
\Phi(\hat\xi)=r^{-\mu} n(z,r) \,.
\label{phitraf2}
\end{equation}

Then Eq.~(\ref{diff-conv2}) becomes
\begin{eqnarray}
&&(\hat\xi^2+1){{d^2\Phi}\over{d\hat\xi^2}}+\left[(1-2\mu)\hat\xi-
{V_0\over{\sqrt{D_{\rm z}\kappa_{\rm r}}}}\right]{{d\Phi}\over{d\hat\xi}}+
\nonumber \\
&&+\mu^2\Phi=-{{Q_0}\over{\kappa_{\rm r} r^{\alpha+\mu-2}}}\delta(z) \,,
\label{Phi}  
\end{eqnarray}
and if  
\begin{equation}
\alpha+\mu-2=-1 \,,
\label{alpha}
\end{equation}
the RHS of Eq.~(\ref{Phi}) can also be cast into self-similar form
\begin{equation}
{{Q_0}\over{\kappa_{\rm r} r^{\alpha+\mu-2}}}\delta(z)={{Q_0}\over
{\sqrt{D_{\rm z}\kappa_{\rm r}}}}\delta(\hat\xi) \,.
\end{equation}

The boundary condition at $\hat\xi=0$ immediately follows from integration 
of Eq.~(\ref{Phi}) for small $\hat\xi$, in which case the equation becomes
\begin{equation}
{d\over{d\hat\xi}}\left({{d\Phi}\over{d\hat\xi}}-{V_0\over
{\sqrt{D_{\rm z}\kappa_{\rm r}}}}\Phi\right)+\mu^2\Phi=-{{Q_0}\over{\sqrt{D_{\rm z}\kappa_{\rm r}}}}
\delta(\hat\xi) \,.
\label{Phi_1}
\end{equation} 
We integrate Eq.~(\ref{Phi_1}) along a box, encircling the Galactic plane 
with $\hat\xi = 0$, from $-\varepsilon$ to $+\varepsilon$, and subsequently
let $\varepsilon \to 0$.
The formal integration near $\hat\xi=0$ gives the boundary condition at 
$\hat\xi=0$,
\begin{equation}
{{d\Phi(0)}\over{d\hat\xi}}-{V_0\over{\sqrt{D_{\rm z}\kappa_{\rm r}}}}\Phi(0)=-{{Q_0}
\over\sqrt{D_{\rm z}\kappa_{\rm r}}} \,.
\end{equation}

The function $\Phi$ is obtained by solving the homogeneous equation
\begin{equation}
(\hat\xi^2+1){{d^2\Phi}\over{d\hat\xi^2}}+[(1-2\mu)\hat\xi-
{V_0\over{\sqrt{D_{\rm z}\kappa_{\rm r}}}}]{{d\Phi}\over{d\hat\xi}}+\mu^2\Phi=0 \,.
\label{Phi_0}
\end{equation}
The solution of Eq.~(\ref{Phi_0}) is found to be 
\begin{eqnarray}
&&\Phi=c_1 \, {}_{2}F_{1}\left(-\mu, -\mu, 1-\mu-\eta;
\frac{1}{2}\left(1\pm i \hat\xi\right)\right) + \nonumber \\
&& c_2 \left[\frac{1}{2}\left(1\pm i  \hat\xi\right)\right]^{\eta+\mu}
\label{hypgeosolution}
\times \\
&& {}_{2}F_{1}\left(\eta, \eta, 1+\mu+\eta;
\frac{1}{2}\left(1\pm i \hat\xi\right)\right) \nonumber \,,
\end{eqnarray}
where
\begin{equation}
\eta={1\over 2}\left(1\mp {{iV_0}\over{\sqrt{D_{\rm z}\kappa_{\rm r}}}}\right)\,.
\end{equation}

Our goal is to study the dependence of the particle density near $z=0$,
in other words we have to determine the value of $\mu$.

 From the asymptotic expansion of hypergeometrical functions at 
$\hat\xi\to\infty$,  
we find that the solution Eq.~(\ref{hypgeosolution}) depends on $\hat\xi$ 
as 
\begin{equation}
\lim_{\xi \to \infty} \Phi(\xi)\propto \hat\xi^\mu \,.
\label{xi_bound}
\end{equation}
This also follows directly from Eq.~(\ref{Phi}), which for large $\hat\xi$ has 
the form
\begin{equation}
\hat\xi^2{{d^2\Phi}\over{d\hat\xi^2}}+(1-2\mu)\hat\xi{{d\Phi}
\over{d\hat\xi}}+\mu^2\Phi=0 \,.
\end{equation}
It is easy to verify that $\Phi$ behaves like a power law in $\hat\xi$, i.e.\
\begin{equation}
\Phi\propto \hat\xi^\lambda \,
\end{equation}
and the characteristic equation
\begin{equation}
\lambda(\lambda-1)+\lambda(1-2\mu)+\mu^2=0
\end{equation}
has just one degenerate solution:
\begin{equation}
\lambda_{1,2}=\mu \,.
\end{equation}

We have already shown that
at large $z$ the distribution function $n$ has the form 
$n \propto 1/z$ (cf.\ Eq.~(\ref{z_infty})). 
On the other hand it follows from Eq.~(\ref{phitraf2}) for $\xi\to\infty$
\begin{equation}
n\propto r^\mu \Phi(\hat\xi) \propto r^\mu \hat\xi^\mu = z^\mu \,
\end{equation}
and therefore $\mu=-1$, thus matching the two solutions obtained 
for different regions in $r-z$-space. It is clear that by the choice of 
the similarity variable (Eq.~(\ref{r_xi})), it is impossible to describe the 
distribution function near the Galactic centre ($r \to 0$), but everywhere 
else in the disk. 

Since close to the sources, $\hat\xi \to 0$, $\Phi \simeq const.$, we derive 
from Eq.~(\ref{phitraf2}) 
\begin{equation}
n(r)\propto {1\over r} \,,
\end{equation}
whereas the source distribution is required to have a radial exponent  
$\alpha = 2$ as follows from Eq.~(\ref{alpha}), and therefore
\begin{equation}
Q(r)\propto {1\over r^2} \,.
\end{equation}

 Thus radially dependent advection
flattens the distribution of CRs in the disk as compared to the source
distribution 
\begin{equation}
f_2(r) \propto r^{-1}, ~~~~~~~~~ f_1 (r) \propto r^{-2}\nonumber\\
\end{equation}
and the ``compensation'' condition, which in this case means a partial 
compensation since $n \propto 1/r$ rather than $n \sim const.$, is realized 
in this case if
\begin{equation}
f_2^2=f_1 \,.
\end{equation}
%
%
   \begin{figure}[thbp]
   \centering
   \mbox{%
   \includegraphics[width=\hsize,clip]{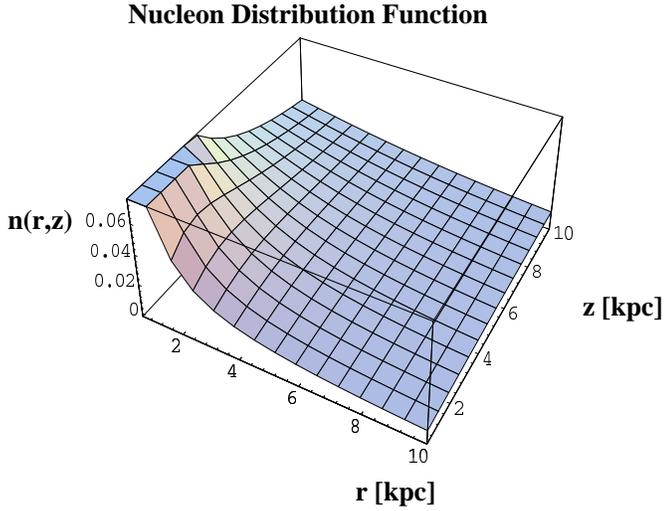}
                   }
      \caption[]{The nucleon distribution function $n(r,z)$ is plotted in
                arbitrary units as a function of Galactocentric radius $r$ and
                height $z$ above the plane. The radial and vertical diffusion
                coefficients are $\kappa_{\rm r} = 3 \times 10^{28} \,
                {\rm cm}^2/{\rm s}$ and $D_{\rm z} = 3 \times 10^{28} \,
                {\rm cm}^2/{\rm s}$, respectively.
                }
         \label{df1}
   \end{figure}
%
%
%
%
   \begin{figure}[thbp]
   \centering
  \includegraphics[width=\hsize,clip]{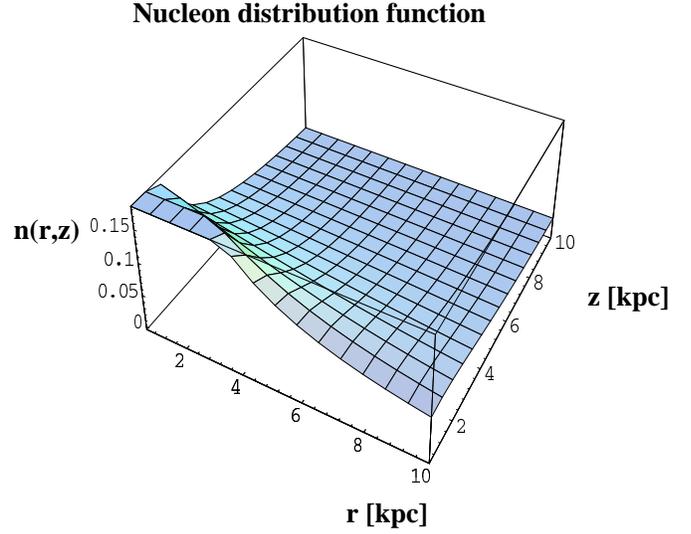}
      \caption[]{Same as Fig.~\ref{df1}, but the radial and vertical diffusion
                coefficients are now different, viz.\ $\kappa_{\rm r} = 3 \times
                10^{29} \, {\rm cm}^2/{\rm s}$ and $D_{\rm z} = 3 \times 10^{28} \,
                {\rm cm}^2/{\rm s}$, respectively. This corresponds to 
		{\em anisotropic} diffusion occurring preferentially in the 
		{\em disk}, where the magnetic field is oriented mainly in 
		radial direction.
                }
         \label{df2}
   \end{figure}
%
%
%
%
   \begin{figure}[thbp]
   \centering
  \includegraphics[width=\hsize,clip]{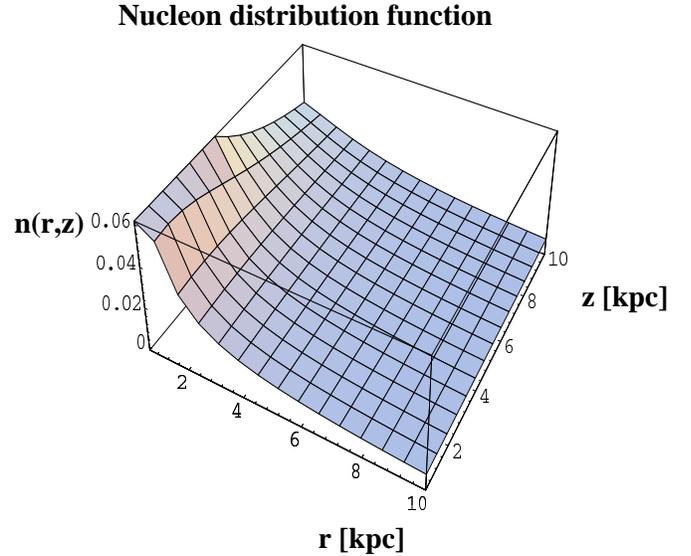}
      \caption[]{Same as Fig.~\ref{df2}, with $\kappa_{\rm r} = 3 \times
                10^{28} \, {\rm cm}^2/{\rm s}$ and $D_{\rm z} = 3 \times 10^{29} \,
                {\rm cm}^2/{\rm s}$, respectively. This corresponds to 
		{\em anisotropic} diffusion occuring preferentially in the 
		{\em halo}, where the magnetic field is oriented mainly in 
		the vertical direction.
                }
         \label{df3}
   \end{figure}
%
In Fig.~\ref{df1} we have plotted the resulting nucleon distribution function, 
which shows a $1/r$-dependence (cf. Eq.~(\ref{diff-conv4})), thus flattening 
out the steeper radially 
dependent source distribution ($\propto 1/r^2$) by a radially varying 
transition boundary. It should be emphasized that such a behaviour is a natural
consequence of the underlying radially varying source distribution itself.
All previous models with $z_{\rm C}(r) = const.$ have ignored the growing
influence of advective transport by
an increase in CR sources and hence local CR pressure.
Of course, it should be kept in mind that $z_{\rm C}$ is also energy
dependent, since from observations of the CR spectrum it is inferred
that $D \propto E^{\delta}$, where the value of $\delta$ depends on 
$z$-variations of the Galactic wind velocity.

Moreover, according to Fig.~\ref{df2}, a stronger radial than vertical mixing
of CRs due to {\em anisotropic} diffusion leads to a further flattening of the 
nucleon distribution function and hence to a weaker diffuse $\gamma$-ray 
gradient. We expect this to happen in the Galactic disk, where the 
geometry of the large scale magnetic field is mainly parallel to the disk. 
In contrast, in the halo a substantial vertical $\vec{B}$-component 
might exist (like e.g.\ in \object{NGC 4631}), resulting in a larger value 
of $D_{\rm z}$ as compared to $\kappa_{\rm r}$. As can be seen from Fig.~\ref{df3} this 
effect is competing to some extent with the also vertically directed 
advection velocity, and therefore the distribution function is similar to 
the one shown in Fig.~\ref{df1}.   

In summary we conclude that the value of the halo extension calculated 
from the gradient data in the framework of the isotropic diffusion model may 
indeed be an artifact.
In the next Sect. we outline a general solution.

\subsection{General solution for axisymmetric con\-vective and an\-iso\-tropic
dif\-fu\-sive CR transport}
\begin{figure}[thbp]
   \centering
   \mbox{%
   \includegraphics[width=0.7\hsize,angle=-90,clip]{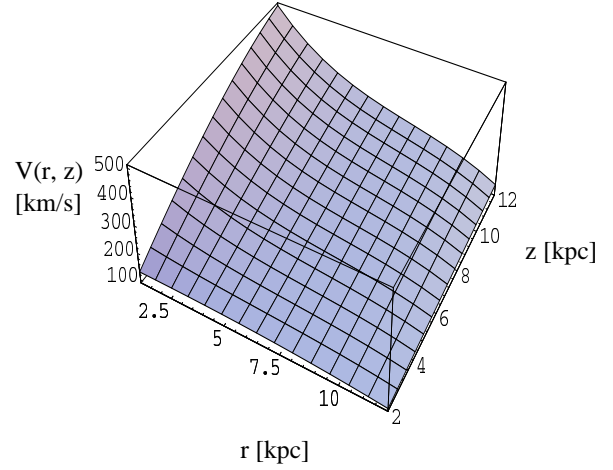} %
                   }
      \caption[]{Galactic wind outflow velocity ($z$-component),  
                $V(r,z) = u_{\rm z}(r,z) + v_{{\rm A}}(r, z)$, as a function of 
                Galactocentric radius $r$ and 
                vertical distance $z$ from the Galactic plane, according to 
                self-consistent Galactic wind calculations for a given 
                SNR distribution (for details see text).
                }
         \label{snd2}
   \end{figure}

We now want to work out a general solution of the stationary two-dimensional 
CR transport equation for nucleons, without any restriction of the radial 
behavior of the source distribution $Q(r,z)$. In the following we use 
$N(r, z, E)$ for the nucleon distribution function in order to avoid 
confusion with the number $n$ for the enumeration of the poles in our 
solutions.  
\begin{eqnarray}
&-& \nabla \left(\underline{\underline D}(\vec{x}, E) \nabla N -
\vec{u}(\vec{x}) N \right) - {\partial \over \partial E} \left(\frac{1}{3}
\nabla\vec{u}(\vec{x}) \, E \, N \right) + \nonumber\\
&+& {\partial \over \partial E} \left({d E \over dt} \, N\right) = 
Q(E, \vec{x}) \,,
\label{2dcon-diff1}
\end{eqnarray}
where the diffusion tensor, $\underline{\underline D}$, in cylindrical 
coordinates with axial symmetry is given by
\begin{equation}
\underline{\underline D} = \left(
\begin{array}{cc}
D_{\rm rr} & D_{\rm rz} \\
D_{\rm zr} & D_{\rm zz}
\end{array}
\right)
= \left(
\begin{array}{cc}
\kappa_{\rm r} & 0 \\
0 & D_{\rm z}\cdot E^{\delta}
\end{array}
\right) \,.
\end{equation}
We are not so much interested in energy dependent than spatially anisotropic
diffusion and advection, and so, for convenience, we set $\delta =0$.
For the nucleon component other than adiabatic losses are negligible
($dE/dt=0$); thus Eq.~(\ref{2dcon-diff1}) in axial symmetry reads
\begin{eqnarray}
&-& D_{\rm z} {\partial^2 N \over \partial z^2} - \kappa_{\rm r}
\left({\partial^2 N \over \partial r^2} + \frac{1}{r} {\partial N \over \partial r}
\right) + 3 {V_0 \over r^2} {\partial \left(N \, z\right) \over \partial z} -
\nonumber \\
&-& {V_0 \over r^2} {\partial \left(N \, E \right) \over \partial E} =
Q_1(r) \, \delta(z) \, E^{-\gamma_0} \,,
\label{2dcon-diff2}
\end{eqnarray}
with
\begin{equation}
Q_1(r) = Q_0 f_1(r) \,,
\end{equation}
and where we have chosen a spatial variation for the velocity field as
\begin{equation}
\vec{u}(\vec{x}) = \left(
\begin{array}{c}
u_{\rm r} \\
u_{\rm z}
\end{array}
\right)
= \left(
\begin{array}{c}
0 \\
3 \frac{V_0}{r^2} z
\end{array}
\right) 
= \left(
\begin{array}{c}
0 \\
3 V_0 f_2(r) z
\end{array}
\right) 
\,,
\label{velvar1}
\end{equation}
and a power law spectrum for injection by the disk
sources. The choice of $\vec{u}(\vec{x})$ is motivated not only by 
simplicity, but also by our numerical calculations (see Figs.~(\ref{snd1})
and (\ref{snd2})). The latter figure shows the outflow velocity as 
observed in a fixed Galactic frame of reference, $V(r,z) = u_{\rm z}(r,z) + 
V_{{\rm A}}(r, z)$. It can be seen that $V(r,z)$ increases roughly 
linearly with $z$ and decreases in a power law fashion with $r$, so that 
Eq.~(\ref{velvar1}) is a good first approximation. The boundary conditions 
for obtaining these results are given in Sect.~\ref{radvar}.
In future calculations, we will
also test other analytical approximations to $V(r, z)$.

It is clear from Eq.~(\ref{2dcon-diff2}) that the effect of anisotropic
diffusion corresponds to stretching or compressing scales in respective 
directions, since in the new variables, $\bar r=r/\sqrt\kappa_{\rm r}$, and,  
$\bar z=z/\sqrt D_{\rm z}$, 
we recover the equation for isotropic diffusion. Obviously, this
may also change the gradient of the cosmic ray density in the disk, but we will 
show that rather strong modifications of the density (including the observed 
uniform distribution of cosmic rays) can be obtained for spatially nonuniform 
advection.

In order to solve Eq.~(\ref{2dcon-diff2}) analytically, we apply the following
transformations
\begin{eqnarray}
N(r,z,E) &=& E^{-\gamma_0} \, S(\xi, \varrho) \,,
\label{vartrans1-1}\\
A &=& \frac{V_0}{\kappa_{\rm r}}\,,
\label{vartrans1-4}\\
B &=& \sqrt{D_{\rm z} \over \kappa_{\rm r}}\,,
\label{vartrans1-5}\\
\gamma &=& \gamma_0+2 \,,
\label{vartrans1-6}\\
\bar r &=& B r \,, \\
\varrho &=& \ln \bar r \qquad \Longrightarrow \bar r = e^{\varrho}.
\label{vartrans1-2}\\
\xi &=& \frac{z}{\bar r} \qquad \Longrightarrow \xi = z e^{-\varrho}\,.
\label{vartrans1-3}
\end{eqnarray}
The differential operators for the variable transformation
$(r,z) \to (\varrho, \xi)$ are given by
\begin{eqnarray}
{\partial \over \partial r} &=& e^{-\varrho} \left({\partial \over \partial\varrho}
- \xi {\partial \over \partial \xi}\right) \,,\\
{\partial \over \partial z} &=& e^{-\varrho} {\partial \over \partial \xi} \,,\\
{\partial^2 \over \partial r^2} &=& e^{- 2 \varrho}\Bigl(
{\xi^2 \, \partial^2 \over \partial \xi^2} - 2 {\xi \, \partial^2 \over \partial \xi
\partial \varrho}
+ 2 {\xi \, \partial \over \partial \xi} + {\partial^2 \over \partial \varrho^2}
- {\partial \over \partial \varrho} \Bigr) \,,\\
{\partial^2 \over \partial z^2} &=& e^{- 2 \varrho} {\partial^2 \over
\partial \xi^2}\,.
\label{vartrans2}
\end{eqnarray}
Using the above transformations,
Eq.~(\ref{2dcon-diff2}) can be cast into
\begin{eqnarray}
(1 &+& \xi^2) {\partial^2 S \over \partial \xi^2} + {\partial^2 S \over \partial
\varrho^2} - 2 \xi {\partial^2 S \over \partial \xi \partial \varrho} +
(1 - 3 A) \, \xi {\partial S \over \partial \xi} \nonumber \\
&-& A \, \gamma \, S = -{Q_1\left(e^{\varrho}\right) \over D_{\rm z}} \,
e^{\varrho} \, \delta(\xi) \,.
\label{2dcon-diff3}
\end{eqnarray}

Next we transform the boundary conditions appropriately:
\begin{eqnarray}
N(r,z=\pm\infty) &=& 0 \quad \longrightarrow \quad S(\varrho, \xi=\pm\infty) 
= 0 \\
N(r=+\infty,z) &=& 0 \quad \longrightarrow \quad S(\varrho=+\infty, \xi) =
0 \\
{{\partial N(r=0,z)}\over{\partial r}} &=& 0 \quad \longrightarrow \quad
\left(e^{-\varrho}{{\partial S}\over
{\partial\varrho}}\right)_{\varrho=-\infty} = 0 \,.
\label{bouco1}
\end{eqnarray}
In order to proceed with the solution of
Eq.~(\ref{2dcon-diff3}), we look for the Green's function $G(\xi,\varrho)$ of
\begin{eqnarray}
(1 &+& \xi^2) {\partial^2 G \over \partial \xi^2} +
{\partial^2 G \over \partial \varrho^2} - 2 \xi {\partial^2 G \over \partial
\xi \partial \varrho} + (1 - 3 A) \, \xi {\partial G \over \partial \xi}
\nonumber \\ &-& A \, \gamma \, G = \delta(\varrho-\varrho^{\prime}) \,
\delta(\xi) \,.
\label{2dcon-diff4}
\end{eqnarray}
with boundary conditions
\begin{eqnarray}
G(\varrho, \xi=\pm\infty) &=& 0 \,,
\label{bouco2}
\\
G(\varrho=\pm\infty, \xi) &=& 0 \,.
\label{bouco3}
\end{eqnarray}
A standard method for
solving PDE's of the above kind is to use integral transforms; since we
consider a steady state problem over all $r$- and $z$-space (or $\xi$ and
$\varrho$, respectively) we use Fourier transforms of the kind
\begin{eqnarray}
G(\xi,\varrho) &=& \int_{-\infty}^{+\infty} {dk \over 2\pi} F(k,\xi)
\exp(-i k \varrho) \,,
\label{foutra1} \\
F(k, \xi) &=& \int_{-\infty}^{+\infty} G(\xi,\varrho) \exp(i k \varrho)
d\varrho \,.
\label{foutra2}
\end{eqnarray}
Inserting the transform Eq.~(\ref{foutra1}) into Eq.~(\ref{2dcon-diff4}), and 
after integrating
by parts we apply the boundary conditions Eqs.~(\ref{bouco2})-(\ref{bouco3}) 
and end up with an ODE in $\xi$,
\begin{eqnarray}
(1 &+& \xi^2)
F^{\prime\prime}(\xi) + \left(2 i k - 3 A + 1\right)\xi F^{\prime}(\xi)
\nonumber \\ &-& \left(k^2 + A \gamma\right) F(\xi) = \delta(\xi) \exp(i k
\varrho^{\prime}) \,.
\label{2dcon-diff5}
\end{eqnarray}
Let us inspect
the associated homogeneous equation first.  Eq.~(\ref{2dcon-diff5}) can be
cast into the hypergeometric differential equation
\begin{equation}
x (x-1) y^{\prime\prime} + \left[(\alpha+\beta+1) x - \tilde\gamma
\right] y^{\prime} + \alpha \beta y = 0 \,,
\label{hypgeo1}
\end{equation}
by
applying the following transformation
\begin{eqnarray}
x &=& p + i q \xi \\
\Rightarrow dx &=& i q \, d\xi \,,
\end{eqnarray}
and
\begin{equation}
x (x-1) = p(p-1) -q^2 \xi^2 + i q \xi (2p - 1) \,,
\end{equation}
from which
we infer $p = 1/2$ and $q=\pm 1/2$.  Matching the 2nd term of
Eq.~(\ref{hypgeo1}) we obtain
\begin{eqnarray}
\alpha+\beta+1 &=& 1-3A+2 i k \,,\\ 
\tilde\gamma &=& \frac{1}{2}(\alpha+\beta+1) \,.
\label{ab1}
\end{eqnarray}
Note that there is no difference between the cases with ``$+$'' and ``$-$''.
The last term in Eq.~(\ref{hypgeo1}) requires
\begin{equation}
\alpha \beta = -(k^2 + \gamma A) \,.
\label{ab2}
\end{equation}
Solving for $\alpha$ and $\beta$ from Eqs.~(\ref{ab1}) and (\ref{ab2}) gives
\begin{eqnarray}
\alpha_{1,2} &=& \frac{1}{2} \left(2 i k - 3 A \pm C\right)\,,
\label{ab3-1}\\
\beta_{1,2} &=& \frac{1}{2} \left(2 i k - 3 A \mp C\right) \,,
\label{ab3-2}
\end{eqnarray}
where $C=\sqrt{A(9A-12 ik + 4 \gamma)}$.
In Eqs.~(\ref{ab3-1}) and (\ref{ab3-2}) $\alpha$ and $\beta$ are complementary
and we can drop the subscripts, using ``+'' for $\alpha$ and ``$-$'' for
$\beta$.

Thus the solution of the homogeneous equation reads
\begin{eqnarray}
F(\xi) &=& c_1 \, {}_{2}F_{1}\left(\alpha, \beta, \tilde\gamma;
\frac{1}{2}\left(1\pm i \xi\right)\right) + \nonumber \\
&& c_2 \left[\frac{1}{2}\left(1\pm i  \xi\right)\right]^{1-\tilde\gamma}
\times \\
&& {}_{2}F_{1}\left(\alpha-\tilde\gamma+1, \beta-\tilde\gamma+1, 2-\tilde\gamma;
\frac{1}{2}\left(1\pm i \xi\right)\right) \nonumber \,,
\label{homsol1}
\end{eqnarray}
where $\tilde\gamma$ is not an integer and $0 < \cal{Re}(1\pm i \xi) < 1$.

Applying the linear transformation
\begin{eqnarray}
{}_{2}F_{1}\left(\alpha, \beta, \tilde\gamma; x\right) &=&
(1-x)^{\tilde\gamma-\alpha-\beta} \times \nonumber \\
&&{}_{2}F_{1}\left(\tilde\gamma-\alpha, \tilde\gamma-\beta, \tilde\gamma; x\right)
\,,
\end{eqnarray}
we finally obtain
\begin{eqnarray}
F(\xi) &=& c_1 \, {}_{2}F_{1}\left(\alpha, \beta, \tilde\gamma;
\frac{1}{2}\left(1\pm i \xi\right)\right) + \nonumber \\
&& c_2 \left[\frac{1}{2}\left(1\pm  i \xi\right)\right]^{1-\tilde\gamma}
\cdot \left[\frac{1}{2}\left(1\mp
i \xi\right)\right]^{\tilde\gamma-\alpha-\beta} \times \nonumber \\
&& {}_{2}F_{1}\left(1-\alpha, 1-\beta, 2-\tilde\gamma;
\frac{1}{2}\left(1\pm i \xi\right)\right) \,,
\nonumber \\
&=& c_1 \, {}_{2}F_{1}\left(\alpha, \beta, \tilde\gamma;
\frac{1}{2}\left(1\pm i  \xi\right)\right) + \nonumber \\
&& c_2 \left[\frac{1}{4}\left(1+ \xi^2 \right)\right]^{1-\tilde\gamma}
\times \nonumber \\
&& {}_{2}F_{1}\left(1-\alpha, 1-\beta, 2-\tilde\gamma;
\frac{1}{2}\left(1\pm i \xi\right)\right) \,,
\label{homsol2}
\end{eqnarray}
where the integration constants $c_1$ and $c_2$ have yet to be determined
from the boundary conditions (see Appendix~\ref{appb}).
 From the Fourier transforms (\ref{foutra1}) and (\ref{foutra2}) we
see that this implies
\begin{equation}
F(\xi = \pm\infty) = 0 \,;
\label{bouco4}
\end{equation}
furthermore we require
\begin{itemize}
\item[(i)] continuity of $F(\xi)$ across $\xi=0$ \,,
\item[(ii)] $F(-\xi) = F(\xi)$ \,.
\end{itemize}
We also note from Eq.~(\ref{homsol2}) that $F(\xi)$ is multivalued in the
complex plane and therefore a branch cut has to be introduced, running along
the imaginary axis from $- i $ to $+ i$.

We now return to the inhomogeneous equation Eq.~(\ref{2dcon-diff5}).
For $\xi \neq 0$, Eq.~(\ref{homsol2}) is a valid solution; in order
to evaluate the solution at $\xi = 0$ we integrate Eq.~(\ref{2dcon-diff5})
in the infinitesimal range $- \varepsilon \leq \xi \leq \varepsilon$, with
$|\varepsilon| \ll 1$ and then take the limit $\varepsilon \to 0$.

Exploiting the properties of the solution $F(\xi)$ outlined above, and
noting that the integration over an odd function vanishes, we obtain
after a little algebra
\begin{equation}
F^{\prime}(0^+) - F^{\prime}(0^-) = \exp(i k \varrho^{\prime}) \,,
\label{jump1}
\end{equation}
which is a jump condition for the derivative at $\xi=0$.
Integrating this result over $\xi$ gives
\begin{equation}
F(0^+) - F(0^-) = 0 \,,
\label{jump2}
\end{equation}
as expected.
With Eqs.~(\ref{bouco4}) and (\ref{jump1}) we can now evaluate the constants
$c_1$ and $c_2$ (see Eqs.~(\ref{coeff-c1}) and (\ref{coeff-c2}), 
Appendix~\ref{appb}).

Manipulating the $\Gamma$-functions with help of the relations
in Appendix~\ref{appb} (\ref{gamrel1} - \ref{gamrel3}) the solution for 
$\xi=0$ reads
\begin{eqnarray}
F(0) &=& \mp \frac{i  \exp(i k \varrho^{\prime})}{2 \pi^2}
\Biggl\{{\Gamma\left(\frac{\alpha}{2}\right)
\Gamma\left(\frac{\beta}{2}\right)
\Gamma\left(\frac{1-\alpha}{2}\right) \Gamma\left(\frac{1-\beta}{2}\right)
\times \over
} \nonumber \\
&& {\times \sin\left(\frac{\pi\beta}{2}\right)
\cos\left(\frac{\pi\alpha}{2}\right) \left[(-1)^{\tilde\gamma}
\cos\left(\frac{\pi\beta}{2}\right) +
\sin\left(\frac{\pi\alpha}{2}\right)\right] \over
(-1)^{\tilde\gamma}
\sin\left(\frac{\pi\beta}{2}\right) +
\cos\left(\frac{\pi\alpha}{2}\right)}\Biggr\} \,.
\nonumber \\
\label{sol-f0}
\end{eqnarray}

\subsection{Solution in the disk}
Using the solution for $F(\xi=0)$ we can now, as a special case, calculate
the full solution in the disk, i.e.\ $N(r, z=0,E)$).
The Green's function is then given by Eq.~(\ref{foutra1})
\begin{equation}
G(\xi=0,\varrho; \varrho^{\prime}) = \int_{-\infty}^{+\infty} {dk \over 2\pi}
F(k,0) \exp(-i k \varrho) \,.
\label{green1} \\
\end{equation}
In our solution (\ref{sol-f0}) we have suppressed the explicit dependence
on $k$, but of course it is implicitly contained in $\alpha(k)$,
$\beta(k)$, and $\tilde\gamma(k)$.

Let $K = i k$, we obtain
\begin{equation}
 G(0;\varrho,
\varrho^{\prime}) = -\frac{i}{2 \pi} \int_{-i \infty}^{+i \infty} \tilde F(K,0)
\exp[-K(\varrho - \varrho^{\prime})] dK \,,
 \end{equation}
 where
 $\tilde F(K,0) = \exp(-K \varrho^{\prime}) F(K,0)$. Then
\begin{equation}
S(\xi, \varrho) =-
\int_{- \infty}^{+ \infty} G(\xi; \varrho,
\varrho^\prime){{Q_1(e^{\varrho^\prime})}\over {D_{\rm z}}}e^{\varrho^\prime}
2\pi e^{\varrho^\prime} d \varrho^\prime
\,.
\label{green3} \\
\end{equation}
Transforming back to the variable $\bar r$ and $r^{\prime}$ 
(Eq.~(\ref{vartrans1-2})):
\begin{equation}
G(0; r, r^{\prime}) = {- i \over 2 \pi}
\int_{-i \infty}^{+i \infty} \tilde F(K,0) \left({\bar r}\over
r^{\prime}\right)^{-K} dK \,,
\label{green2} \\
\end{equation}
and
\begin{equation}
S(\xi, r) =-
\int_{0}^{+ \infty} G(\xi; r,
r^\prime){{Q_1(r^\prime)}\over {D_{\rm z}}}2\pi (r^\prime)^2{{d\varrho^\prime}\over {d
r^\prime}}d r^\prime
 \,.
\label{green4} \\
\end{equation}

The particle distribution in the disk is then given by
\begin{eqnarray}
&& N(r, z=0, E) = {- i \over 2 \pi D_{\rm z}} E^{-\gamma_0}  \times \nonumber \\ 
&& \times \int_{0}^{\infty}
Q_1(r^{\prime})2\pi r' dr'
\int_{-i \infty}^{+i \infty} \tilde F(K,0) \left({\bar r}\over
r^{\prime}\right)^{-K} dK \,
\end{eqnarray}
and using the expression (\ref{sol-f0}) for $F(K,0) = F(\xi=0)$, we obtain
\begin{eqnarray}
&& N(r, z=0, E) = \nonumber\\
&&\mp \frac{1}{4 \pi D_{\rm z}} E^{-\gamma_0} \int_{0}^{\infty}
Q_1(r^{\prime}) \,
2 \pi r^\prime dr^{\prime} \times\nonumber\\
&\times&\int_{-i \infty}^{+i \infty}
{{\Gamma\left(\frac{\alpha}{2}\right)
 \Gamma\left(\frac{1-\beta}{2}\right)}
 \over
{\Gamma\left(\frac{1+\alpha}{2}\right)
 \Gamma\left(1-\frac{\beta}{2}\right)}}\times \\
&\times&
\left\{(-1)^{\tilde\gamma} \cos\left(\frac{\pi\beta}{2}\right) +
\sin\left(\frac{\pi\alpha}{2}\right)
\over
(-1)^{\tilde\gamma} \sin\left(\frac{\pi\beta}{2}\right) +
\cos\left(\frac{\pi\alpha}{2}\right)\right\} \,
\left(r\over r^{\prime}\right)^{-K} dK 
\nonumber \,.
\end{eqnarray}
With the help of the Green's function from Appendix~\ref{appc} 
(cf.\ Eqs.~(\ref{greenf1}) and (\ref{greenf2}) we obtain
\begin{eqnarray}
&& N(r, z=0, E) =
\mp \frac{1}{4 \pi D_{\rm z}} E^{-\gamma_0} \times
\nonumber \\
&&
\times \Biggl\{\int_{r}^{+\infty}
Q_1(r^{\prime}) 2 \pi r^\prime dr^\prime \int_{C_\alpha}^{}
{{\Gamma\left(\frac{\alpha}{2}\right)
 \Gamma\left(\frac{1-\beta}{2}\right)}
 \over
{\Gamma\left(\frac{1+\alpha}{2}\right)
 \Gamma\left(1-\frac{\beta}{2}\right)}}
\nonumber \\
&&\times
\left\{(-1)^{\tilde\gamma} \cos\left(\frac{\pi\beta}{2}\right) +
\sin\left(\frac{\pi\alpha}{2}\right)
\over
(-1)^{\tilde\gamma} \sin\left(\frac{\pi\beta}{2}\right) +
\cos\left(\frac{\pi\alpha}{2}\right)\right\} \,
\left({\bar r}\over r^{\prime}\right)^{-K} {{dK}\over{d\alpha}}d\alpha +\,
\nonumber \\
&& +\int_{0}^{r}
Q_1(r^{\prime}) 2 \pi r^\prime dr^\prime \int_{C_\beta}^{}
{{\Gamma\left(\frac{\alpha}{2}\right)
 \Gamma\left(\frac{1-\beta}{2}\right)}
 \over
{\Gamma\left(\frac{1+\alpha}{2}\right)
 \Gamma\left(1-\frac{\beta}{2}\right)}}\times \\
&&\times
\left\{(-1)^{\tilde\gamma} \cos\left(\frac{\pi\beta}{2}\right) +
\sin\left(\frac{\pi\alpha}{2}\right)
\over
(-1)^{\tilde\gamma} \sin\left(\frac{\pi\beta}{2}\right) +
\cos\left(\frac{\pi\alpha}{2}\right)\right\} \,
\left({\bar r}\over r^{\prime}\right)^{-K} {{dK}\over{d\beta}}d\beta\Biggr\} \,
\nonumber \,,
\end{eqnarray}
where the integration contours $C_\alpha$ and $C_\beta$ correspond to $r^\prime
>r$ and $r^\prime < r$, respectively.
The only poles to consider for $A \gg 1$ are (see Fig.~\ref{cont1})
for $n = 0,1,2, ...$ 
\begin{eqnarray}
\alpha(K_n) &=& - 2n \quad \Longrightarrow K_n = -2 n \pm \sqrt{A(\gamma + 6
n)} \,,
\end{eqnarray}
where the sign "$-$" should be chosen for $K_n$.
The poles $\alpha(K_n)$ are inside the contour $C_\alpha$ 
(see Appendix~\ref{appc}).

Then we get from the theorem of residues (the residues of Gamma functions
$\Gamma (n)$ at $n=0,-1,-2,...$ are given by ${{(-1)^n}\over {n!}}$)
\begin{eqnarray}
 && N(r, z=0, E) =\nonumber\\
&&\mp \frac{1}{4 \pi D_{\rm z}}
E^{-\gamma_0} \Biggl\{\int_{\bar r}^{+\infty} Q_1(r^{\prime}) 2 \pi r^\prime
dr^\prime\nonumber\\
&& \sum_{n=0}^{\infty} {{(-1)^n
 \Gamma\left(\frac{1-\beta_n}{2}\right)}
 \over
{n! \Gamma\left(\frac{1-2n}{2}\right)
 \Gamma\left(1-\frac{\beta_n}{2}\right)}}\times
 \\
&\times&
\left\{(-1)^{\tilde\gamma_n} \cos\left(\frac{\pi\beta_n}{2}\right)
\over
(-1)^{\tilde\gamma_n} \sin\left(\frac{\pi\beta_n}{2}\right)+ (-1)^n
\right\} \,
\left({\bar r}\over r^{\prime}\right)^{-K_n} {{dK_n}\over {d\alpha}}
\nonumber \,.
\label{dissol1}
\end{eqnarray}
Here
\begin{eqnarray}
\beta_n &=& -2n - 3A - 2\sqrt{A(\gamma + 6n)} \\
\tilde\gamma_n &=& {1\over 2}\left(1-4n - 3A -
2\sqrt{A(\gamma+6n)}\right) \\
K_n &=& - 2n - \sqrt{A(\gamma + 6n)} \\
{{dK_n}\over{d\alpha}} &=& 1 +
{{3A}\over{2\sqrt{A(\gamma+6n)}}} \,. 
\end{eqnarray}
Although Eq.~(\ref{dissol1}) looks rather awkward, this solution for 
$A \gg 1$ exhibits simple asymptotic dependencies at different radii.  
If the CR sources occupy
a limited volume of the disk, bounded by a radius $a$, $Q(r^\prime )= Q_0   
r'^{-q} \Theta (a-r^\prime)$, then for $r>a$ the function $N$ is zero, 
since the
integral ${\int_{r}^{+\infty}}=0$. This means that advection can remove 
particles
from the plane $z=0$ so fast that none remain in the plane 
at radii $r>a$ (see Appendix~\ref{appc}).

Inside the source region ($r\ll a$), $N(r)$  can be 
written as
\begin{equation}
N(r)\propto r^{\sqrt{A\gamma}}{\int_r^{+\infty}}{r'}^{(1-q-\sqrt{A\gamma})} 
dr \,,
\end{equation}
so that for strong advection
$A \gg 1$ the  integral is determined by the lower limit
{\em independent} of the source distribution. Then we have
\begin{equation}
N(r)\propto r^{2-q} \,,
\end{equation}
and the CR density is almost constant in the disk region
if $q \approx 2$. We have therefore demonstrated that in the case of strong 
and spatially nonuniform advection, the CR density in the disk exhibits only 
a weak gradient in spite of strong radial variations of the source 
distribution. Such a behaviour may be responsible for the mild gradient 
observed in the diffuse $\gamma$-ray data.

In the diffusion dominated case ($A \ll 1$) the values of the integrals are 
determined by the source
distribution. Here we should take into account the integral over $C_\beta$, 
which is not zero (see Appendix~\ref{appc}).
Far away from the disk, i.e.\  $r \gg a$, the distribution function is 
determined
by the only pole, $\beta = 1$, which is inside the contour $C_\beta$ (see
Fig.~\ref{cont5}). Therefore independent of the source distribution, we have
\begin{equation}
N(r)\propto r^{-1-\sqrt{A(\gamma-3)}} \propto r^{-1} \,,
\end{equation}
just what is expected for pure diffusive particle propagation.

Therefore we conclude from these analyses that a more or less uniform CR 
distribution in the disk can be expected, if advection is strong, 
i.e.\ $A\gg 1$.

The ``compensation'' equation in this case reads
\begin{equation}
f_2(r)=f_1(r) \,.
\end{equation}

The interesting conclusion from the solution of the general
three-dimensional model is that we can reach a quasi-uniform CR distribution 
especially when the boundary $z_{\rm C}$ is {\em closest} to the Galactic 
plane in regions where the source density and hence advection is strongest 
(see also Fig.~\ref{snd3}), i.e.\ when $A\gg 1$. 
This is in sharp contrast to the results of the pure diffusion model, 
because only an extremely extended galactic halo can produce 
a more or less smooth CR distribution and thus a mild diffuse $\gamma$-ray 
gradient.

\section{Source contributions to the $\gamma$-ray background}
\label{sourcecon}
The examples of the last Sect. substantiate our argument that transport
effects should be responsible for all but eliminating the $\gamma$-ray
gradient in the GeV energy range. This ignores possible contributions to
the $\gamma$-ray flux from the sources themselves. In fact, SNRs as CR
sources are expected to accumulate the accelerated CRs before
releasing them into the ISM at the end of their lifetime. In nuclear
collisions with thermal gas atoms or (for electrons) with photons inside
the remnant, this high density of energetic particles will produce a
strong $\gamma$-ray intensity that contributes a ``diffuse'' background if
these sources are unresolved. In fact the ensemble of Galactic SNRs,
expected to be strongly concentrated in the disk, will constitute such a
background with typically less than about 10 such sources along a line of
sight in the disk. They will be unresolved in GeV $\gamma$-rays, except for
a few well-known nearby objects (Berezhko \& V\"olk 2000); also their
collective flux contributes less than $10$\% to the overall $\gamma$-ray
flux. However, we know that the CR source spectrum is much harder $\propto
E^{-(2.1 {\rm to} 2.2)}$ than the diffuse Galactic CR energy spectrum 
($\propto
E^{-2.7}$), and therefore the source contribution will equal the truly
diffuse $\gamma$-ray flux at about 100 GeV and become even dominant at
about $10^3$ GeV (= 1 TeV). 

If the sources collectively dominate in the form of an unresolved
``diffuse'' background at very high $\gamma$-ray energies, then it should
be possible to observe the average CR source distribution and even the
average CR source spectrum (!) along the Galactic disk in the TeV range.
Present observations at TeV energies have not yet been able to detect this
``diffuse'' background. However, the experimental upper limits and the
predicted ``diffuse'' flux are so close to each other (Aharonian et al.  
2001) that an instrumental increase by an order of magnitude should lead
to a detection. The next generation of TeV instruments like H.E.S.S.,
CANGAROO, MAGIC or Veritas, that has this sensitivity, will come on line soon.  
The comparison of such a detected TeV - distribution with the radial shape
of the source distribution from radio studies would constitute a stringent
empirical test on our theoretical arguments about the transport effects on
the truly diffuse Galactic CR component. It also shows the basic
importance of $\gamma$-ray surveys over a large range in energies and
radial distance.

\section{Discussion and Conclusions}
\label{discon}
In the detailed analysis of the previous Sections it was shown 
that a physically reasonable CR transport model should include the 
nonlinear dynamical effects of the 
Galactic distribution of the CR sources themselves. Taking into 
account a simple radial dependence of the source power as suggested 
by radio observations of supernova remnants and pulsars, and thus 
averaging over all azimuthal angles, 
leads already to a radically different spatial behaviour of the 
CR distribution function. In all cases we have studied so far, there  
is a clear tendency to smooth out peaks due to enhanced particle injection 
by the sources in the ensuing radial particle distribution. This can be 
attributed to {\em CR advection in an outflow} aided or even caused by the CRs 
themselves. In other words, the importance of CR advection is proportional 
to the CR power of the underlying sources. As a consequence, e.g.\ the CR 
escape time, which is usually inferred from secondary radioactive 
isotopes, need no longer be a globally averaged quantity, but could 
vary strongly from place to place in the Galaxy.  
Therefore, the usual assumption that parameters of the CR flux measured near 
Earth are representative for the global processes of CR propagation in 
the Galaxy may be a misleading oversimplification. 

 It is common practice to determine from the intensities of different 
CR components near Earth the average
diffusion coefficient in the Galaxy, the velocity of advection, the 
height of the CR halo in the direction perpendicular to the Galactic plane, 
and the CR injection spectrum, just to name the most important ones. 
Based on this hypothesis the nearly uniform radial 
CR distribution, derived from the measurement of diffuse Galactic 
$\gamma$-rays, can be reproduced only
if there exists thorough spatial mixing of CRs in the framework of an extended 
halo (if CR diffusion is isotropic). Hence in such a case {\em local} and 
{\em global} properties of CRs do not differ from each other. 
However, the inferred halo height from chemical composition 
($z_{\rm C} = 2 - 4$ kpc; see e.g., Bloemen et al.  1991, 1993; Webber et al. 1993; 
Lukasiak et al. 1994) is clearly inconsistent with the value derived from 
the interpretation of the $\gamma$-ray data ($z_{\rm C} = 15 - 20$ kpc; cf.\ Dogiel 
\& Uryson 1988, Bloemen et al.  1993) 
within the framework of an isotropic diffusion model (see Appendix~\ref{appa}). 
We therefore conclude that the halo size derived
from CR nuclear data reflects only a local value near Earth, and 
the huge halo extension derived previously from $\gamma$-ray data may be an 
artifact, since it relies on the validity of global values for locally obtained 
CR data. This conclusion is supported by our numerical galactic wind 
simulations, which show that the vertical distance of the diffusion-advection 
transition boundary from the Galactic plane, is inversely proportional to 
the CR source power and not spatially constant as been previously assumed.  
 
Radio observations of external galaxies indicate a large-scale magnetic field 
geometry, which is mainly parallel to the major axis in the disk, and if 
a halo field exists, it is parallel to the minor axis.
Therefore we expect that CR diffusion is in general {\em anisotropic}, with
a radial diffusion coefficient $\kappa_{\rm r}$ in the disk, which is much 
larger than diffusion in the perpendicular direction, $D_{\rm z}$, and vice versa 
in the halo. In this case the initially inhomogeneous CR 
distribution, due to a radially varying source distribution in the disk, is 
smeared out, whereas in the halo the dominant diffusion component $D_{\rm z}$ 
can be superposed by a strong advection velocity, which may determine 
the spatial particle distribution.  

It would be desirable to have a high enough spatial resolution and 
photon statistics in the future to observe the radial distribution 
of diffuse $\gamma$-rays above 100 MeV in nearby edge-on galaxies, 
such as NGC253. However, it seems unlikely that both space-borne and 
ground-based $\gamma$-ray observatories will satisfy this requirement  
in the near future. Thus the only direct observation of the CR source 
distribution in the Galaxy will be possible with next generation 
TeV instruments like H.E.S.S. 

\begin{acknowledgements}
DB acknowledges support from the
Deutsche Forschungsgemeinschaft (DFG) by a Heisenberg fellowship.
He thanks the Max-Planck-Institut f\"ur Kernphysik in Heidelberg, the
Max-Planck-Institut f\"ur extraterrestrische Physik in Garching, 
and the Department of Astrophysical Sciences of Princeton University, 
where this research has been carried out, for support and hospitality. 
DB thanks Russell Kulsrud for many interesting discussions. 

VAD acknowledges financial support  from the Alexander von Humboldt-Stiftung
which was very essential for these collaborative researches. 
This work was prepared during his visit to Max-Planck-Institut f\"ur 
extraterrestrische Physik (Garching) and he is grateful to his colleagues from
this institute for helpful and fruitful discussions. 
The final version of the paper was partly done at the Institute of Space and 
Astronautical Science. VAD thanks his colleagues from the institute and 
especially Prof.\ H.\ Inoue for their warm hospitality.

\end{acknowledgements}

\appendix
\section{The halo size from  local characteristics of the CR flux}
\label{appa} 
  
The flux of CRs observed near Earth (locally) contains information about 
regions from where these high energy particle have been produced, as well as
about the location of the boundary from 
where these particle escape into the intergalactic medium and do not return 
to the galactic disk. 
This means that in principle we can infer the size of 
the diffusion region simply from local CR properties. 
Let us denote the thickness of this region by $z_{\rm esc}$. 
We will demonstrate how to derive this quantity  
from a set of simplified equations, describing densities of stable and 
radioactive secondary nuclei generated by primary CRs in the gaseous disk. 

For a one-dimensional
diffusion model the system of equations for the density of stable, $n_{\rm s}$, and
radioactive, $n_{\rm r}$, nuclei is given by
\begin{equation}
D{{{d^2}{n_{\rm s}}}\over
{dz^2}}=-{N_{\rm g}} v {\sigma_{\rm sp}} {z_{\rm d}} {n_{\rm p}^0} \delta (z)
\nonumber\,,
\end{equation}
and
\begin{equation}
D{{{d^2}{n_{\rm r}}}\over{d{z^2}}}-{{n_{\rm r}}\over \tau}=- {N_{\rm g}} v {\sigma_{\rm rp}}
{z_{\rm d}} {n_{\rm p}^0}\delta (z)
\nonumber\,.
\end{equation}
The distribution function of the nuclei in the disk is not very sensitive 
to the boundary conditions. Therefore we analyze here the simplest case, 
in which  at $z_{\rm esc}$ it is required that 
\begin{equation}
{n_{\rm s}},{n_{\rm r}}=0
\nonumber\,;
\end{equation}
here $\sigma_{\rm sp}$ and $\sigma_{\rm rp}$ are the spallation cross sections 
of stable and radioactive isotopes, respectively. 
We assume that processes
of nuclear spallations take place in the gaseous disk with 
half-thickness $z_{\rm d}$, which is much less than the halo scale 
height $z_{\rm esc}$.

Here $z$ is the coordinate perpendicular to the Galactic plane; the 
gas is supposed to be concentrated in the Galactic disk with ${z_{\rm d}} \ll {z_{\rm esc}}$, 
where the gas density equals $N_{\rm g}$, and $n_{\rm p}^0$ is
the density of primary cosmic rays in the disk $(z=0)$.
Then
\begin{equation}
{n_{\rm s}}={N_{\rm g}} v {\sigma_{\rm sp}} {n_{\rm p}^0} {{z_{\rm d}}\over D}\left(z_{\rm esc} -z \right) \,,
\end{equation}
and correspondingly 
\begin{equation}
{n_{\rm r}}={N_{\rm g}} v {\sigma_{\rm rp}} {n_{\rm p}^0}{{{z_{\rm d}}{\sqrt{D\tau}}}\over D}
{{{\sinh}({{{z_{\rm esc}}-z}\over {\sqrt{D\tau}}})}\over {{\cosh}({{z_{\rm esc}}\over
{\sqrt{D\tau}}})}}
\nonumber\,,
\end{equation}
and thus the ratio ${{n_{\rm s}}\over {n_{\rm r}}}$ at $z=0$ is 
\begin{equation}
f={{n_{\rm r}}\over {n_{\rm s}}}= {{\sigma_{\rm rp}}\over {\sigma_{\rm rs}}}{{\sqrt
{D\tau}}\over {z_{\rm h}}}{\tanh {{z_{\rm esc}}\over{\sqrt{D\tau}}}}
\nonumber\,.
\end{equation}
The CR life-time in the region of diffusion propagation is determined from
\begin{equation}
\tau_{\rm esc} \sim {{z_{\rm esc}^2}\over D}
\nonumber\,,
\end{equation}
and for $\tau \ll \tau_{\rm esc}$ we obtain
\begin{equation}
f={n_{\rm r} \over n_{\rm s}}\simeq {\sigma_{\rm rp}\over \sigma_{\rm sp}}
\sqrt{\tau\over \tau_{\rm esc}} = {\sigma_{\rm rp} \over \sigma_{\rm sp}} 
{\sqrt{D\tau } \over z_{\rm esc}}  \, 
\label{ratio}\,.
\end{equation}
We see that this ratio depends on the two unknown parameters
$D$ and $z_{\rm esc}$. Therefore they cannot be determined
independently. To eliminate one parameter, we need additional 
observational constraints, for instance
the average thickness of matter traversed by CR nuclei in
the Galaxy during their lifetime, given by the expression
\begin{equation}
x=M_{\rm p} N_{\rm g} v {z_{\rm d}\over z_{\rm esc}}{\tau_{\rm esc}}\,,
\end{equation}
where $v$ is the velocity of the nuclei and $M_{\rm p}$ is the proton rest mass.

Combining this with Eq.~(\ref{ratio}) we obtain
\begin{equation}
x\simeq {{M_{\rm p} z_{\rm d} N_{\rm g} v\tau}\over{f^2 z_{\rm esc}}} \left({{\sigma_{\rm rp}}\over 
{\sigma_{\rm sp}}}\right)^2\,,
\end{equation}
which gives $z_{\rm esc}\simeq 1$  kpc, for $z_{\rm d} N_{\rm g}\simeq 4\cdot
10^{20}$cm$^{-2}$, $x\simeq 12 - 14$ g cm$^{-2}$ (Ferrando et al. 1991,
Ferrando 1993), $\tau = 2.2\cdot 10^6$ years and $f\simeq 0.25$ (see e.g.,
Simpson \& Garcia-Mu\~noz 1988). Here we took $\sigma_{\rm rp}\sim\sigma_{\rm sp}$ 
for the estimate.

If we include the convection term into the kinetic equation 
the analysis becomes more complicated, but the result is almost the same. 
If the velocity depends on $z$ as
\begin{equation}
V(z)=3 V_0 z \,,
\end{equation}
we obtain for the ratio $f$ (see Bloemen et al. 1993)
\begin{equation}
f \simeq \sqrt{3 V_0 \tau} \,,
\label{f}
\end{equation}
which requires $V_0\simeq 10$ km s$^{-1}$ kpc$^{-1}$. The grammage in this 
case is
\begin{equation}
x\sim M_{\rm p} N_{\rm g} v {{z_{\rm d} z_{\rm esc}}\over{D(E)}}\,,
\label{x}
\end{equation}
where $z_{\rm esc}$ is the coordinate of the surface above the Galactic plane 
from where CRs are sucked away by the Galactic wind (i.e.\ the transition 
boundary between diffusion and advection). 
\begin{equation}
z_{\rm esc} \sim \sqrt{{D(E)}\over V_0}\,.
\end{equation}

Combining  Eqs.~(\ref{f}) - (\ref{x}) we also obtain for  
($\sigma_{\rm rp}\sim\sigma_{\rm sp}$)
\begin{equation}
x\simeq {{M_{\rm p} z_{\rm d} N_{\rm g} v\tau}\over{f^2 z_{\rm esc}}}
\end{equation}
and $z_{\rm esc}$ is also about 1 kpc. 

This value estimated from the chemical composition of CRs near Earth is 
usually defined as the size of the CR halo. On the other hand the 
halo size can be derived from the global $\gamma$-ray emissivity distribution 
in the Galactic plane. The value of $z_{\rm esc}$ then determines the 
size of the CR mixing volume, which is estimated to be at least 10 kpc.  
If the diffusion model would describe the CR distribution correctly 
and satisfactorily in the Galaxy, 
the estimates of $z_{\rm esc}$ obtained from {\em local} and {\em global} 
data would be similar.
However, the sizes of the CR propagation region differ from each other by 
about one order of magnitude, ruling out pure diffusion as the sole CR 
transport process. 

\section{Determination of the integration constants $c1$ and $c2$}
\label{appb}

Using the transformation formula
\begin{eqnarray}
{}_{2}F_{1}\left(\alpha, \beta, \tilde\gamma; x\right) &=& {\Gamma(\tilde\gamma)
\Gamma(\beta-\alpha) \over \Gamma(\beta) \Gamma(\tilde\gamma-\alpha)}
(-x)^{-\alpha} \times \nonumber \\
& & {}_{2}F_{1}\left(\alpha, 1-\tilde\gamma+\alpha, 1-\beta+\alpha;
\frac{1}{x}\right) + \nonumber \\
& & {\Gamma(\tilde\gamma) \Gamma(\alpha-\beta) \over \Gamma(\alpha)
\Gamma(\tilde\gamma-\beta)} (-x)^{-\beta} \times \nonumber \\
& & {}_{2}F_{1}\left(\beta, 1-\tilde\gamma+\beta, 1-\alpha+\beta;
\frac{1}{x}\right) \,,
\label{hypgeo2}
\end{eqnarray}
observing that $x= \frac{1}{2}\left(1\pm {i \over B} \xi\right)$,
and the limit (which of course will also hold upon interchanging $\alpha$ and
$\beta$ due to their symmetry, see Eqs.~(\ref{ab1}) and (\ref{ab2}))
\begin{eqnarray}
&&\lim_{\xi \to \pm\infty} {}_{2}F_{1}\left(\alpha, 1-\tilde\gamma+\alpha,
1-\beta+\alpha; \frac{1}{\frac{1}{2}\left(1\pm {i \over B} \xi\right)}\right)
\nonumber \\
&=& 1 \,,
\label{bouco22}
\end{eqnarray}
we obtain from the condition (\ref{bouco4}) for Eq.~(\ref{homsol2}) 
after some algebra
\begin{eqnarray}
&&\lim_{\xi \to \pm\infty} \Biggl\{e^{i \pi \alpha}
\left[\frac{1}{2}\left(1\pm {i \over B} \xi\right)\right]^{-\alpha}
\times \nonumber \\
&&{}_{2}F_{1}\left(\alpha, 1-\tilde\gamma+\alpha, 1-\beta+\alpha;
\frac{1}{x}\right)
\Biggl[c_1 {\Gamma(\tilde\gamma) \Gamma(\beta-\alpha) \over \Gamma(\beta)
\Gamma(\tilde\gamma-\alpha)} + \nonumber \\
&&c_2 \, e^{- i \pi(\tilde\gamma-1)}
{\Gamma(2-\tilde\gamma) \Gamma(\beta-\alpha) \over \Gamma(\beta -
\tilde\gamma + 1) \Gamma(1-\alpha)} \Biggr] \nonumber \\
&+& e^{i \pi \beta}
\left[\frac{1}{2}\left(1\pm {i \over B} \xi\right)\right]^{-\beta}
\times \nonumber \\
&&{}_{2}F_{1}\left(\beta, 1-\tilde\gamma+\beta, 1-\alpha+\beta;
\frac{1}{x}\right)
\Biggl[c_1 {\Gamma(\tilde\gamma) \Gamma(\alpha-\beta) \over \Gamma(\alpha)
\Gamma(\tilde\gamma-\beta)} + \nonumber \\
&&c_2 \, e^{- i \pi(\tilde\gamma-1)}
{\Gamma(2-\tilde\gamma) \Gamma(\alpha-\beta) \over \Gamma(\alpha -
\tilde\gamma + 1) \Gamma(1-\beta)} \Biggr]\Biggr\} = 0 \,.
\label{bouco33}
\end{eqnarray}
Observing that $Re(a)>0$ and $Re(b)<0$, Eq.~(\ref{bouco33}) can only be satisfied
in general, if
\begin{equation}
c_1 = (-1)^{\tilde\gamma} {\Gamma(2-\tilde\gamma) \Gamma(\alpha)
\over \Gamma(\tilde\gamma) \Gamma(1-\beta)} \, c_2 \,.
\label{bouco44}
\end{equation}
Next we use condition (\ref{jump1}) to fix $c_1$ and $c_2$.
Making use of the relations
\begin{eqnarray}
&&{d \over dx}\,{}_{2}F_{1}\left(\alpha, \beta, \tilde\gamma; x\right) =
\nonumber \\
&&\qquad{\alpha \, \beta \over \tilde\gamma} 
{}_{2}F_{1}\left(\alpha+1, \beta+1, \tilde\gamma+1; x\right) \,, \\
&&{}_{2}F_{1}\left(\alpha+1, \beta+1, \tilde\gamma+1; \frac{1}{2}\right) =
{4 \sqrt{\pi} \tilde\gamma \over \alpha \, \beta}{\Gamma(\tilde\gamma) \over
\Gamma(\frac{\alpha}{2}) \Gamma(\frac{\beta}{2})} \,,\\
&&{}_{2}F_{1}\left(2-\alpha, 2-\beta, 3-\tilde\gamma; \frac{1}{2}\right)
\nonumber \\
&&\qquad= {4 \sqrt{\pi} \, \Gamma(3-\tilde\gamma) \over (1-\alpha) (1-\beta)
\Gamma(\frac{1-\alpha}{2}) \Gamma(\frac{1-\beta}{2})} \,,
\end{eqnarray}
we obtain
\begin{eqnarray}
\lim_{\xi \to 0}F^{\prime}(\xi) & = & \exp(i k \varrho^{\prime}) \nonumber \\
& = & \pm 4 \pi {i \over 2 B} \Biggl\{c_1 {\Gamma(\tilde\gamma) \over
\Gamma(\frac{\alpha}{2}) \Gamma(\frac{\beta}{2})}  \nonumber \\
& \quad & +c_2 4^{\tilde\gamma -1)}
{\Gamma(2-\tilde\gamma) \over \Gamma(\frac{1-\alpha}{2})
\Gamma(\frac{1-\beta}{2})}\Biggr\} \,.
\label{bouco55}
\end{eqnarray}
Inserting Eq.~(\ref{bouco44}) into Eq.~(\ref{bouco55}) we end up
after some tedious algebra and using the following relations
\begin{eqnarray}
\Gamma(1+z) &=& z \Gamma(z)
\label{gamrel1} \\
\Gamma(z) \Gamma(1-z) &=& \frac{\pi}{\sin(\pi z)}
\label{gamrel2} \\
\Gamma(2 z) &=& \frac{1}{\sqrt{\pi}} 2^{2 z -1} \Gamma(z)
\Gamma\left(z+\frac{1}{2}\right)
\label{gamrel3} \,,
\end{eqnarray}
with
\begin{eqnarray}
c_1 &=& \mp \frac{i B \exp(i k \varrho^{\prime})}{2 \sqrt{\pi}}
{(-1)^{\tilde\gamma} \Gamma(\frac{\alpha}{2}) \Gamma(\frac{\beta}{2})
\sin\left(\frac{\pi\beta}{2}\right) \over
\Gamma(\tilde\gamma)\left[(-1)^{\tilde\gamma}
\sin\left(\frac{\pi\beta}{2}\right) +
\cos\left(\frac{\pi\alpha}{2}\right)\right]}
\,, \nonumber\\
\label{coeff-c1}
\end{eqnarray}
and
\begin{eqnarray}
c_2 &=& \mp \frac{i B \exp(i k \varrho^{\prime})}{\sqrt{\pi}}{2^{-(\alpha+\beta)}
\cos\left(\frac{\pi\alpha}{2}\right) \Gamma(\frac{1-\alpha}{2})
\Gamma(\frac{1-\beta}{2}) \over \Gamma(2-\tilde\gamma)
\left[(-1)^{\tilde\gamma}
\sin\left(\frac{\pi\beta}{2}\right) +
\cos\left(\frac{\pi\alpha}{2}\right)\right]}
\,. \nonumber\\ 
\label{coeff-c2}
\end{eqnarray}

\section{Determination of integration contours for the Green's 
function $G(\xi, \bar r, r^\prime)$}
\label{appc}

The Fourier transform
\begin{eqnarray}
F(k, \xi) &=& \int_{-\infty}^{+\infty} G(\xi,\varrho) \exp(i k \varrho)
d\varrho \,,
\end{eqnarray}
of the Green's function of Eq.~(\ref{2dcon-diff4}) is shown in 
Eq.~(\ref{foutra2}) with
the constants $c_1$ and $c_2$ taken from  Eqs.~(\ref{coeff-c1}) and 
(\ref{coeff-c2}). For the
special case of the solution in the Galactic plane ($\xi=0$) Eq.~(\ref{homsol2})
becomes
\begin{eqnarray}
F(\xi=0) &=& \mp \frac{i  \exp(i k \varrho^{\prime})}{2}
\Biggl\{{{\Gamma\left(\frac{\alpha}{2}\right)
 \Gamma\left(\frac{1-\beta}{2}\right)}
 \over
{\Gamma\left(\frac{1+\alpha}{2}\right)
 \Gamma\left(1-\frac{\beta}{2}\right)}}\times
\nonumber \\
&& \times{{ \left[(-1)^{\tilde\gamma}
\cos\left(\frac{\pi\beta}{2}\right) +
\sin\left(\frac{\pi\alpha}{2}\right)\right]} \over
{\left[(-1)^{\tilde\gamma}
\sin\left(\frac{\pi\beta}{2}\right) +
\cos\left(\frac{\pi\alpha}{2}\right)\right]}}\Biggr\}
 \,.
\end{eqnarray}
Then the Green's function for $\xi=0$ reads
\begin{eqnarray}
G(\xi=0,\varrho) &=& \int_{-\infty}^{+\infty} {dk \over 2\pi} F(k,\xi=0)
\exp(-i k \varrho) \nonumber \,.
\end{eqnarray}
If $K=ik$, then the transform for the variable $r$ is
\begin{equation}
G(0; \bar r, r^{\prime}) = {- i \over 2 \pi} \int_{-i \infty}^{+i \infty}
\tilde F(K,0) \left({\bar r}\over r^{\prime}\right)^{-K} dK \nonumber \,,
\label{greenf3}
\end{equation}
which is determined by the residues of the poles of the $\Gamma$-functions
\begin{equation}
\alpha(K_n) = -2n
\label{alpha1}
\end{equation}
and
\begin{equation}
\beta(K_m) = 2m + 1 \,,
\label{beta1}
\end{equation}
and by the residues of the poles determined from
\begin{equation}
(-1)^{\tilde\gamma}\sin\left({{\pi\beta}\over
2}\right)+\cos\left({{\pi\alpha}\over2}\right)=0
\,.
\label{pole}
\end{equation}
Since $\tilde\gamma=\alpha+\beta+1$ this equation can be written in the form
\begin{eqnarray}
&&\exp\left\lbrack i\pi\left(\alpha +{\beta\over 2}-{1\over
2}\right)\right\rbrack -
\exp\left\lbrack i\pi\left(\alpha -{\beta\over 2}-{1\over 2}
\right)\right\rbrack +
\nonumber \\
&&\exp\left\lbrack i\pi\left(-\beta -1+{\alpha\over 2}\right)\right\rbrack +
\exp\left\lbrack i\pi\left(-\beta -1-{\alpha\over 2} \right)\right\rbrack
\nonumber \\
&=& 0
\,,
\end{eqnarray}
yielding
\begin{equation}
\beta=-\alpha-1+4n \,.
\end{equation}
However, these values of $\beta$ also make the numerator vanish, i.e.\ 
\begin{equation}
(-1)^{\tilde\gamma}\cos\left({{\pi\beta}\over
2}\right)+\sin\left({{\pi\alpha}\over2}\right)=0
\,,
\end{equation}
because all we are doing is to interchange $\alpha$ and $\beta$.
Thus $\beta=-\alpha -1+4n$ does not give any poles.

\begin{figure}[thbp]
   \includegraphics[width=0.9\hsize,angle=90,bbllx=535pt,bblly=0pt, %
                bburx=0pt,bbury=575pt,clip]{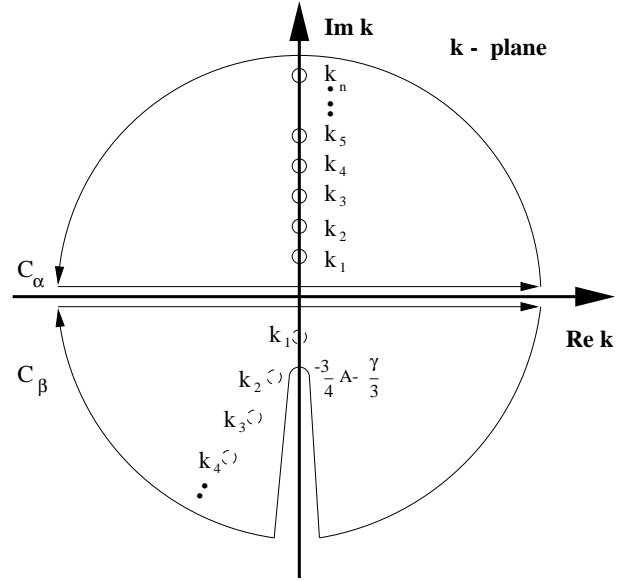}
   \caption[]{
The contours of integration $C_\alpha$ and $C_\beta$ in Eq.~(\ref{greenf1})
                for $\bar r<r'$ and $\bar r>r'$  in the $k$-plane for
$A\gg 1$.
To show these contours in the $K=ik$ plane the figure should be
rotated by an angle of $\pi /2$. The dashed circles represent poles of the 
second kind, the solid circles poles of the first kind (see text).
}
\label{cont1}
\end{figure}

\begin{figure}[thbp]
     \includegraphics[width=0.9\hsize,angle=90,bbllx=535pt,bblly=0pt, %
                bburx=0pt,bbury=575pt,clip]{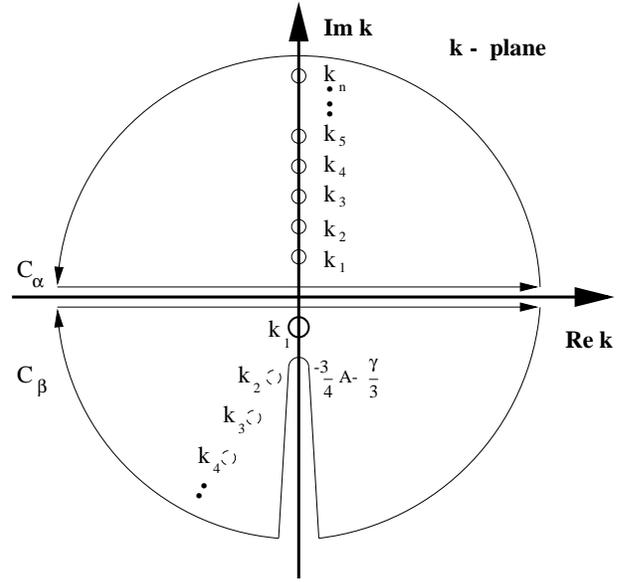}
      \caption[]{
The contours of integration $C_\alpha$ and $C_\beta$ in Eq.~(\ref{greenf1})
                for $\bar r<r'$ and $\bar r>r'$  in the $k$-plane for
$A\ll 1$.
To show these contours in the $K=ik$ plane the figure should be
rotated by an angle of $\pi /2$. The dashed circles represent poles of the 
second kind, the solid circles poles of the first kind (see text).
}
\label{cont2}
\end{figure}

We see from Eq.~(\ref{greenf3}) that the contour of integration should be 
enclosed in the
semiplane ${\rm Re}(K>0)$ for $\bar r>r^\prime$, i.e.\ the contour $C_\alpha$ 
in Figs.~\ref{cont1} and \ref{cont2}.
In this case the integration over the outer part of the closed contour is 
zero and
the function G equals the residues of the sum inside the closed contour only. 
For
$\bar r<r^\prime$ the contour is enclosed in the outer part of the 
semiplane ${\rm Re}(K<0)$, i.e.\ the contour $C_\beta$ in 
Figs.~\ref{cont1} and \ref{cont2}.
The locations of the poles in the $K$-plane are determined from a one-to-one 
map of the contours onto the $\alpha$- and $\beta$-planes and 
subsequently Eqs.~(\ref{alpha1}) and (\ref{beta1}) are applied.

The Green's function reads
\begin{eqnarray}
G(\xi&=&0; r, r^{\prime}) = {- i \over 2 \pi}
\Biggl\{\Theta (r^\prime-r)\int_{C_\alpha} \tilde F(K_\alpha,\xi) \left(\bar
r\over r^{\prime}\right)^{-K_\alpha} \times\nonumber \\
&\times&{{dK_\alpha}\over {d\alpha}}d\alpha +
\Theta(r - r^\prime)\int_{C_\beta} \tilde F(K_\beta,\xi) \left(\bar r\over
r^{\prime}\right)^{-K_\beta} \times\nonumber \\ 
&\times& {{dK_\beta}\over {d\beta}}d\beta \Biggr\}
\,,
\label{greenf1}
\end{eqnarray}
and applying the theorem of residues we obtain
\begin{eqnarray}
&&G(\xi=0,r,r') = \nonumber\\
&&\frac{-i}{2\pi}\Biggl\{\Theta(r'-r)\Bigl[\sum\mbox{Res}
\left(\Gamma\left({\alpha\over 2}\right)\right)\Bigr]
{{\Gamma\left(\frac{1-\beta}{2}\right)} \over
{\Gamma\left(\frac{1+\alpha}{2}\right)
 \Gamma\left(1-\frac{\beta}{2}\right)}}\times
\nonumber \\
&&\times
\left\{(-1)^{\tilde\gamma} \cos\left(\frac{\pi\beta}{2}\right) +
\sin\left(\frac{\pi\alpha}{2}\right)
\over
(-1)^{\tilde\gamma} \sin\left(\frac{\pi\beta}{2}\right) +
\cos\left(\frac{\pi\alpha}{2}\right)\right\} \,
\left(r\over r^{\prime}\right)^{-K}+  \,
\nonumber \, \\
&&+\Theta(r'-r) \Bigl[\sum \mbox{Res} \left(
\Gamma\left(\frac{1-\beta}{2}\right) \right)\Bigr]
{{\Gamma\left({\alpha\over 2}\right)}
\over
{\Gamma\left(\frac{1+\alpha}{2}\right)
 \Gamma\left(1-\frac{\beta}{2}\right)}}\times
\nonumber \\
&&\times
\left\{(-1)^{\tilde\gamma} \cos\left(\frac{\pi\beta}{2}\right) +
\sin\left(\frac{\pi\alpha}{2}\right)
\over
(-1)^{\tilde\gamma} \sin\left(\frac{\pi\beta}{2}\right) +
\cos\left(\frac{\pi\alpha}{2}\right)\right\}
\left(r\over r^{\prime}\right)^{-K}\Biggr\}  \,,
\label{greenf2}
\end{eqnarray}
where  $\alpha, \beta, \tilde\gamma$ and $K$ are determined by their values at
the poles of the Gamma functions. The residues of the function $\Gamma (n)$ 
have the value ${{(-1)^n}\over {n!}}$ at $n= 0, -1, -2, ....$.

The contours for $\bar r>r^\prime$ and $\bar r<r^\prime$ in the $k$-plane are
shown for $A \gg 1$ in Fig.~\ref{cont1}, and for $A \ll 1$ in Fig.~\ref{cont2}.

%


To determine the locations of the poles of the $\Gamma$-functions 
$\alpha({K_n})$ and
$\beta({K_m})$ in the $K$-plane we should map a unique array of
$K$ values one-to-one onto corresponding arrays of $\alpha$ and $\beta$ 
values. In this way we
define the contours of ${\rm Re}(K)>0$ and ${\rm Re}(K)<0$ in  
$\alpha$ and $\beta$ planes.

We start with $r>r^\prime$.
   \begin{figure}[thbp]
     \includegraphics[width=0.9\hsize,angle=90,bbllx=540pt,bblly=0pt, %
                bburx=0pt,bbury=425pt,clip]{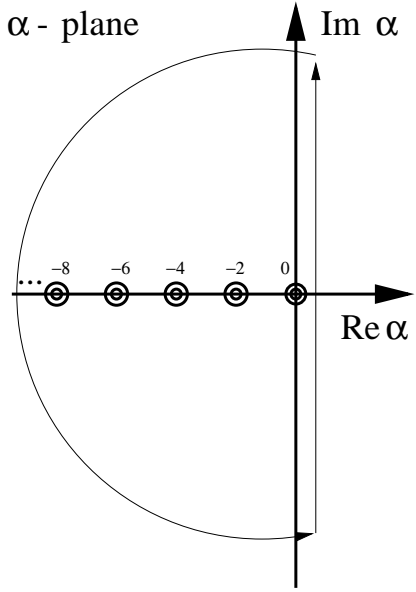}
      \caption[]{
The contour of integration $C_\alpha$
                for $\bar r<r'$   in the $\alpha$-plane.
The solid circles represent poles inside the contour $C_\alpha$.
}
\label{cont3}
\end{figure}
%
%
%
We find from $\alpha(K=0)$ (see Fig.~\ref{cont3}) that
the contour $C_\alpha$ intersects the abscissa at the points:
\begin{equation}
-\infty~~~~~~~~\mbox{and}~~~~~~~~
-{{3A}\over2}+{1\over 2}\sqrt{A\left(9A+4\gamma\right)}
\end{equation}
Therefore all poles $\alpha_n$ are inside the contour
$C_\alpha$. On the other hand the area encircled by $C_\alpha$ does not contain
poles $\beta_m$. We see that poles $\alpha_n$ mapped onto the $k$-plane are
transformed to the negative part of the real axis (see Figs.~\ref{cont1},  
\ref{cont2} and \ref{cont3}).

\begin{figure}[thbp]
     \includegraphics[width=0.9\hsize,angle=90,bbllx=535pt,bblly=115pt, %
                bburx=0pt,bbury=585pt,clip]{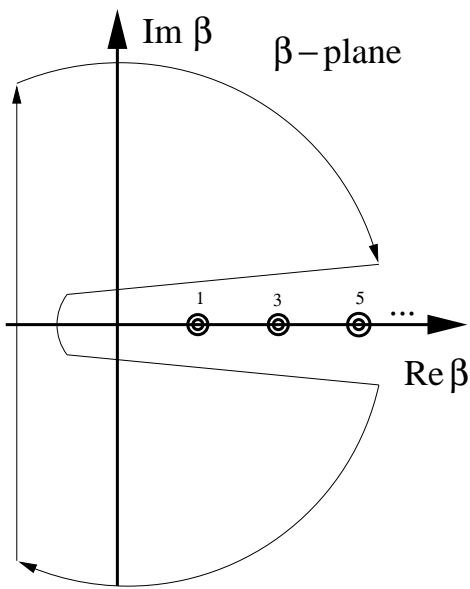}
      \caption[]{
The contour of integration $C_\beta$ and the poles $\beta_m$
                for $\bar r>r'$   in the $\beta$-plane for $A\gg 1$.
 }
\label{cont4}
\end{figure}

\begin{figure}[thbp]
    \includegraphics[width=0.9\hsize,angle=90,bbllx=525pt,bblly=0pt, %
                bburx=0pt,bbury=455pt,clip]{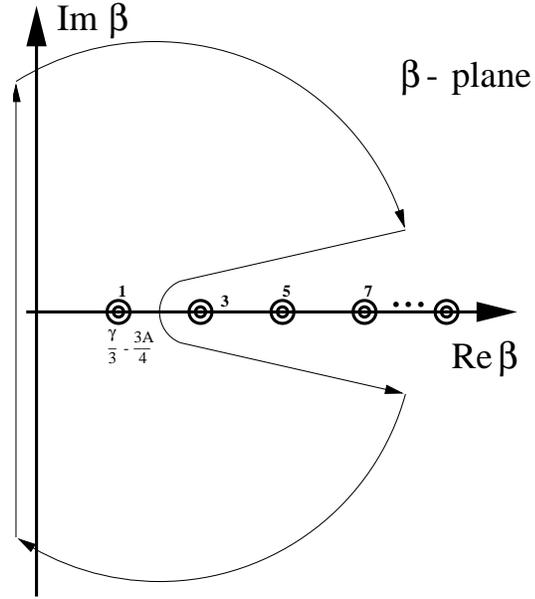}
      \caption[]{
The contour of integration $C_\beta$
                for $\bar r>r'$   in the $\beta$-plane for $A\ll 1$.
}
         \label{cont5}
   \end{figure}

Since there is a branch point in the $k$-plane at
\begin{equation}
k=-i\left({{3A}\over 4}+{\gamma\over 3}\right)
\end{equation}
the contour of integration $C_\beta$ ($r>r'$)  has a cut (see
Figs.~\ref{cont1} and \ref{cont2}).
Its reflection in the $\beta$-plane is shown in
Figs.~\ref{cont4} and \ref{cont5}.  We see that
the contour $C_\beta$ intersects the abscissa axis at the points
\begin{equation}
-{{3A}\over 2}-{1\over 2}\sqrt{A(9A+4\gamma)}
\end{equation}
and
\begin{equation}
-{{3A}\over 4}+{\gamma\over 3}
\,.\end{equation}
The value of the latter is negative for $A \gg 1$ and positive for 
$A\ll 1$ ($\gamma \simeq 4$). Therefore 
only if $A\ll 1$ the pole $\beta=1$ is inside the contour
$C_\beta$ (see Fig.~\ref{cont5}). Correspondingly, there is only one pole 
inside the contour $C_\beta$ in the $k$-plane (cf.\ Fig.~\ref{cont2}).
In the opposite case, $A \gg 1$, all poles are outside this area
(see Fig.~\ref{cont4}) and then the integral over $C_\beta$ equals zero. 
This means
that the poles $\beta_m$ mapped onto the $k$-plane are of the second kind 
(their phases are larger than $2\pi$, see Fig.~\ref{cont1}). The reason, 
is that in the case
of strong advection and the velocity increasing towards the center 
as $r^{-2}$,
advection removes cosmic rays from the central part of the disk so fast that
these particles do not reach its outer parts. Therefore the contribution of the
integral over $C_\beta$ (the central part of the disk) is zero and the Green's 
function is determined by the integral over $C_\alpha$ only.

%
%
%
%
%
%
%
%
We conclude this Sect. by deriving the transformations between 
$K$-, $\alpha$- and $\beta$-planes, and by deriving the Green's functions 
both for strong and weak advection.
 From the condition of one-to-one correspondence we obtain
for the poles in the $K$-plane
\begin{eqnarray}
\alpha(K_n) &=& -2n \quad \Longrightarrow K_n = -2n -
\sqrt{A(\gamma + 6 n)}\,.
\end{eqnarray}
\begin{eqnarray}
\beta(K_m) = 2m + 1 \, \Longrightarrow
K_m &=& 2m+1+ \nonumber\\
&&\sqrt{A(\gamma - 3(2m+1))}\,.
\end{eqnarray}
Since
\begin{eqnarray}
\alpha(K_n) &=& -2 n
\label{gam-pol3a}\\
\beta(K_m) &=& 2 m+1
\label{gam-pol3b}\,,
\end{eqnarray}
for $n,m = 0,1,2, ...$ we have
\begin{eqnarray}
\beta_n &=& -2n - 3A - 2\sqrt{A(\gamma + 6n)} \\
\tilde\gamma_n &=& {1\over 2}\left(1-4n - 3A -
2\sqrt{A(\gamma+6n)}\right)\\
K_n &=& - 2n - \sqrt{A(\gamma + 6n)}\\
{{dK_n}\over{d\alpha}} &=& 1 +
{{3A}\over{2\sqrt{A(\gamma+6n)}}}\\
\alpha_m &=& 2m + 1 - 3A + 2
\sqrt{A(\gamma - 3(2m+1))} \\
\tilde\gamma_m &=& 
\frac{1}{2}\Bigl(2(2m+1)-3A+1+ \nonumber\\
&&2\sqrt{A(\gamma-3(2m+1))}\Bigr) \\
K_m &=& 2m+1+\sqrt{A(\gamma-3(2m+1))}\\
{{dK_m}\over{d\beta}} &=& 1 - {{3A}\over{2\sqrt{A(\gamma-3(2m+1))}}}
\end{eqnarray}
and for $A \gg 1$ (cf.\ Eq.~(\ref{greenf1})):
\begin{eqnarray}
&& G(\xi=0, r, r') = \nonumber\\
&&
 \frac{-i}{2 \pi}
 \Biggl\{\Theta (r'-r)
\sum_{n=0}^{\infty} {{(-1)^n
 \Gamma\left(\frac{1-\beta_n}{2}\right)}
 \over
{n!\Gamma\left(\frac{1-2n}{2}\right)
 \Gamma\left(1-\frac{\beta_n}{2}\right)}}\times
\\
&\times&
\left\{(-1)^{\tilde\gamma_n} \cos\left(\frac{\pi\beta_n}{2}\right)
\over
(-1)^{\tilde\gamma_n} \sin\left(\frac{\pi\beta_n}{2}\right)+ (-1)^n
\right\} \,
\left({\bar r}\over r^{\prime}\right)^{-K_n} {{dK_n}\over {d\alpha}}
\Biggr\} \nonumber 
\,.
\end{eqnarray}
For $A \ll 1$ the Green's function reads
\begin{eqnarray}
&& G(\xi=0, r, r') = \nonumber \\
&&
 \frac{-i}{2 \pi}
 \Biggl\{\Theta (r'-r)
\sum_{n=0}^{\infty} {{(-1)^n
 \Gamma\left(\frac{1-\beta_n}{2}\right)}
 \over
{n!\Gamma\left(\frac{1-2n}{2}\right)
 \Gamma\left(1-\frac{\beta_n}{2}\right)}}\times
\nonumber \\
&\times&
\left\{(-1)^{\tilde\gamma_n} \cos\left(\frac{\pi\beta_n}{2}\right)
\over
(-1)^{\tilde\gamma_n} \sin\left(\frac{\pi\beta_n}{2}\right)+ (-1)^n
\right\} \,
\left({\bar r}\over r^{\prime}\right)^{-K_n} {{dK_n}\over {d\alpha}}+\nonumber\\
&&+
\Theta(r-r')
{{
 \Gamma\left(\frac{\alpha_\beta}{2}\right)}
 \over
{\Gamma\left(\frac{1}{2}\right)
 \Gamma\left(1+\frac{\alpha_\beta}{2}\right)}}\times
\nonumber \\
&\times&
\left\{
\sin\left(\frac{\pi\alpha_\beta}{2}\right)
\over
(-1)^{\tilde\gamma_\beta}  +
\cos\left(\frac{\pi\alpha_\beta}{2}\right)
\right\}
\left({\bar r}\over r^{\prime}\right)^{-K_\beta} {{dK_\beta}\over
{d\beta}}\Biggr\} \,,
\end{eqnarray}
where
\begin{eqnarray}
\beta &=& 1 \\
\alpha_\beta &=& 1-3A+2\sqrt{A(\gamma-3)} \\
\tilde\gamma_\beta &=& {1\over
2}\left(2-3A+1+2\sqrt{A(\gamma-3)}\right)\\
K_\beta &=& 1+\sqrt{A(\gamma-3)}\\
{{dK_m}\over{d\beta}} &=& 1 - {{3A}\over{2\sqrt{A(\gamma-3)}}} \,.
\end{eqnarray}


\begin{thebibliography}{}


\bibitem[2001]{aharon01}
Aharonian, F.A. et al., 2001, A\&A submitted

\bibitem[1999]{aharon99}
Aharonian, F.A., \& Atoyan, A.M. 1999, A\&A 362, 937

\bibitem[1977]{axetal77} 
Axford, W.I., Leer, E., \& Skadron, G. 1977, in: Proceedings of 15th Int.\ 
Cosmic Ray Conf.\ (Plovdiv), Bulgarian Academy of Sciences, 2, 273

\bibitem[1980]{bah:80}
Bakhareva, M.V., \& Smirnova, V.V. 1980, Geomagnetism and Aeronomy 20, 5.

\bibitem[1978a]{be78a}
Bell, A.R. 1978a, MNRAS 182, 147

\bibitem[1979b]{be78b}
Bell, A.R. 1978b, MNRAS 182, 443

\bibitem[1994]{berezhko:94}
Berezhko, E.G., Yelshin, V.K., \& Ksenofontov, L.T. 1994, APh 2, 215

\bibitem[2000]{berezhko:00}
Berezhko, E.G., \& V\"olk, H.J. 2000, ApJ 540, 923

\bibitem[1990]{berezin:90}
Berezinsky, V.S., Bulanov, S.V., Dogiel, V.A., Ginzburg, V.L., \& Ptuskin, V.S.
1990,
{\em Astrophysics of Cosmic Rays}, (ed. V.L.Ginzburg), North Holland.

\bibitem[1978]{blos78}
Blandford, R.D., \& Ostriker, J.P. 1978, ApJ 221, L29

\bibitem[1991]{bloemen:91}
Bloemen, J.B.G.M., Dogiel, V.A., Dorman, V.L., \& Ptuskin, V.S. 1991,  
Izv.AN SSSR, ser.\ fizich., 55, 2052

\bibitem[1993]{bloemen:93}
Bloemen, J.B.G.M., Dogiel, V.A., Dorman, V.L., \& Ptuskin, V.S. 1993,
A\&A, 267,  372

\bibitem[1994]{b94}
Breitschwerdt D., 1994, Habilitation-Thesis (University Heidelberg), 158 p.

\bibitem[1987]{bmv87}
Breitschwerdt D., McKenzie J.F., \& V\"olk H.J. 1987, in: Interstellar 
Magnetic Fields, eds.\ R.\ Beck and R.\ Gr\"ave, Springer, 131


\bibitem[1991]{breit:91}
Breitschwerdt D., McKenzie J.F., \& V\"olk H.J. 1991, A\&A, 245, 79

\bibitem[1993]{breit:93}
Breitschwerdt D., McKenzie J.F., \& V\"olk H.J. 1993,  A\&A, 269, 54

\bibitem[1994]{breit:94}
Breitschwerdt, D.,  \& Schmutzler, T. 1994, Nature 371, 774

\bibitem[1999]{bs99}
Breitschwerdt, D.,  \& Schmutzler, T. 1999, A\&A 347, 650

\bibitem[1992]{bf92}
Bykov, A.M., \&  Fleishman, G.D. 1992, MNRAS 255, 269

\bibitem[1996]{cb96}
Case, G.L., \& Bhattacharya, D. 1996, AAS 120, 437

\bibitem[1998]{cb98}
Case, G.L., \& Bhattacharya, D. 1998, ApJ 504, 761

\bibitem[1992]{dett92}
Dettmar, R.-J. 1992, Fund.\ Cos.\ Phys.\ 15, 143

\bibitem[1996]{dig:96}
Digel,S.W., Grenier, I.A., Heithausen, A., Hunter, S.D., 
\& Thaddeus, P. 1996, ApJ, 463, 609.

\bibitem[1997]{dogiel:97}
Dogiel, V. A., \& Sch\"onfelder V. 1997,  in: Radiophysics and Quantum
Electronics, 40, 57


\bibitem[1988]{dogiel:88}
Dogiel, V. A., \& Uryson, A. V. 1988, A\&A 197,  335

\bibitem[1994]{dogiel:94}
Dogiel, V.A., Gurevich, A.V., \& Zybin, K.P. 1994, A\&A 281, 937

\bibitem[1999]{dragi:99}
Dragicevich, P.M., Blair, D.G., \& Burman, R.R. 1999, MNRAS 302, 693

\bibitem[1989]{dru89} 
Drury, L.O'C., Markiewicz, W.J., \& V\"olk, H.J. 1989, A\&A 225, 179

\bibitem[1989]{evans:89}
Evans, R., van den Bergh, \& S., McClure, R.D. 1989, ApJ 345, 752

\bibitem[1993]{fer:93}
Ferrando P., 1993, Invited, Rapporteur and Highlight Papers, (eds.
Leahy D.A., Hicks R.B., Vankatesan D.), World Scientific,
23rd ICRC, Calgary

\bibitem[1991]{fer:91}
Ferrando P., Lal N., McDonald F. B., \& Webber W. R.
 1991, A\&A 247, 163

\bibitem{fi91}
Fichtner, H., Neutsch, W., Fahr, H.J., \& Schlickeiser, R. 1991, ApJ 371, 98

\bibitem[1964]{ginzburg:64}
Ginzburg, V.L., Syrovatskii, S.I. 1964,  in: The Origin of Cosmic Rays,
Pergamon Press

\bibitem[1999]{hig:99}
Higbie P.R., Peterson J.D., Rockstroh J.M., \& Webber W.R. 1999, 
Proc. of the 26th ICRC, 4, 251 

\bibitem[1987]{hum87}
Hummel, E., Lesch, H., Wielebinski, R., \& Schlickeiser, R. 1988, 
A\&A 197, L29. 

\bibitem[1975]{ip75}
Ipavich, F.M. 1975, ApJ 196, 107

\bibitem[1974]{kod:74}
Kodeira, K. 1974, PASP 26, 266

\bibitem[1977]{kr77} Krymsky, G.F. 1977, Dokl.\ Akad.\ Nauk.\ SSSR 234, 1306

\bibitem[1969]{kul:69}
Kulsrud, R.M., \& Pearce, W.D. 1969, ApJ 156, 445

\bibitem[1989]{lx89}
Leahy, D.A., \& Xinji, W. 1989, PASP 101, 607

\bibitem{li85}
Li, Z.W. 1985, in: Proceedings of Bejing Normal University, 4, 65

\bibitem[1994]{luk:94}
Lukasiak, A., Ferrando, P., McDonald, F.B., \& Webber, W.R. 1994, ApJ 423, 426

\bibitem[1982]{mv82}
McKenzie, J.F., \& V\"olk, H.J. 1982, A\&A 116, 191

\bibitem[1999]{nt99}
Neuh\"auser, \& R., Tr\"umper, J. 1999, A\&A 343, 151

\bibitem[1996]{nor:96}
Normandeau, M., Taylor, A.R., \& Dewdney, P.E. 1996, Nature 380, 687

\bibitem[1997]{pt:97}
Ptuskin, V.S., V\"olk, H.J., Zirakashvili, V.N., \& Breitschwerdt, D. 1997, 
A\&A 321, 434

\bibitem[1976]{se:76} Seiradakis, J.H. 1976, in: The Structure and Content of the 
Galaxy and Galactic Gamma Rays (NASA CP-002; Washington: Government Printing 
Office), p.~265 

\bibitem[1988]{SG88}
Simpson, J.A., \& Garcia-Mu\~noz, M. 1988, SSR 46, 205

\bibitem[1977]{stecker:77}
Stecker, F.W., \& Jones, F.C. 1977, ApJ 217, 843

\bibitem[1996]{strong:96}
Strong, A. W., Benett, K., \& Bloemen, H., et al. 1996,
AAS (3rd COMPTON Symp., M\"unchen, 1995) 120, 381

\bibitem[1988]{strong:88}
Strong, A.W., Bloemen, J.B.G.M.,  Dame, T.M., et al. 
1988, A\&A 207, 1

\bibitem[1996]{str:96}
Strong, A.W., \& Mattox, J.R., 1996, A\&A 308, 21

\bibitem[1997]{strong:98}
Strong, A.W., \& Moskalenko, I.V. 1998, ApJ 509, 212

\bibitem[2000]{sm00}
Strong, A.W., Moskalenko, I.V., \& Reimer, O. 2000, ApJ 537, 763


\bibitem[1985]{top:85}
Toptygin, I.N. 1985, Cosmic Rays in Interplanetary Magnetic Fields,
Reidel, Amsterdam

\bibitem[2000]{tuell00}
T\"ullmann, R., Dettmar, R.-J., Soida, M., Urbanik, M., \& Rossa, J. 2000,
A\&A 364, L36

\bibitem[1990]{vB:90}
van den Bergh, S. 1990, AJ 99, 843

\bibitem[1991]{vB:91}
van den Bergh, S. 1991, in: Supernovae, ed.\ Woosley, S., Springer, 
New York, p.~711 

\bibitem[2000]{voelk:00}
V\"olk, H.J. 2000, in AIP Conf. Proc. 515, GeV-TeV Gamma Ray 
Astrophysics Workshop: Towards a Major Atmospheric Cherenkov Detector VI, 
ed. B.L. Dingus, M.H. Salamon \& D.B. Kieda (New York: AIP), 281


\bibitem[1995]{wa:95} 
Wang, Q.D., Walterbos, R.A.M., Steakley, M.F., Norman, 
C.A., \& Braun, R. 1995, ApJ 439, 176

\bibitem[1992]{webber:92}
Webber, W.R., Lee, M.A., \& Gupta, M. 1992, ApJ 390, 96


\bibitem[1974]{wentzel:74}
Wentzel, D.G. 1974, ARA\&A 12, 71

\bibitem[1996]{zir:96}
Zirakashvili, V.N., Breitschwerdt, D.,  Ptuskin, V.S., \& V\"olk, H.J. 1996,
A\&A 311, 113



\end{thebibliography}
\end{document}